\documentclass{aa}  
\authorrunning{Zielinski et al.}
\usepackage{graphicx}
\usepackage{color}
\usepackage{upgreek}
\usepackage{footnote}
\makesavenoteenv{tabular}
\usepackage[normalem]{ulem}
\usepackage{txfonts}

\begin{document}

   \title{Magnetic field structure of OMC-3 in the far infrared revealed by SOFIA/HAWC+}
   \titlerunning{Magnetic field structure of OMC-3}


   \author{
   N. Zielinski and S. Wolf
          }
    \authorrunning{N. Zielinski}

   \institute{University of Kiel, Institute of Theoretical Physics and Astrophysics, Leibnizstrasse 15, 24118 Kiel, Germany\\
              \email{nzielinski@astrophysik.uni-kiel.de}}

   \date{Received / Accepted }

 
  \abstract
{  
 We report the SOFIA/HAWC+ band D (154\,$\mu$m) and E (214\,$\mu$m) polarimetric observations of the filamentary structure OMC-3 that is part of the Orion molecular cloud.
 The polarization pattern is uniform for both bands and parallel to the filament structure. The polarization degree decreases toward regions with high intensity for both bands, revealing a so called "polarization hole." We identified an optical depth effect in which polarized emission and extinction act as counteracting mechanisms as a potential contributor to this phenomenon. 
   Assuming that the detected polarization is caused by the emission of magnetically aligned non-spherical dust grains, the inferred magnetic field is uniform and oriented perpendicular to the filament. The magnetic field strength derived from the polarization patterns at 154\,$\mu$m and 214\,$\mu$m amounts to 202\,$\mu$G and 261\,$\mu$G, respectively.
  
The derived magnetic field direction is consistent with that derived from previous polarimetric observations in the far infrared and submillimeter (submm) wavelength range.
 Investigating the far-infrared polarization spectrum derived from the SOFIA/HAWC+ observations, we do not find a clear correlation between the polarization spectrum  and cloud properties, namely, the column density, $N(H_2),$ and  temperature, $T$.
}

   \maketitle
%

\section{Introduction}
Magnetic fields in astrophysical objects are ubiquitous and can be found on both small and large scales. However, the role of magnetic fields  in various physical processes, in particular, star formation, is matter of ongoing debate. 
For instance, magnetic fields are considered as a mechanism which slows down the contraction of star-forming regions and filaments, thus, providing a possible explanation for the low star formation rates that have been observed \citep{VanLoo2015, Federrath2015}.
 Polarimetric oberservations of the thermal reemission radiation of magnetically aligned non-spherical dust grains can be used to derive the magnetic field structure and strength in star-forming environments. \\
It is assumed that radiative torque (RAT) alignment is the underlying process for aligning the dust grains. Here, elongated dust grains spin up and align with their longer axis perpendicular to  the magnetic field lines due to the Barnett effect \citep{Barnett1915, Lazarian2007, LazarianHoang2007, Hoang2009} in the presence of an anisotropic radiation field (e.g., a central star embedded in a circumstellar envelope). \\
A broad range of attempts has been undertaken to detect this polarized radiation, such as with JCMT/SCUBA-2 \citep{Holland2013}, the Planck satellite \citep{PlanckCollaboration2011}, and ALMA. In recent years, HAWC+ \citep{Harper2019} aboard the Stratospheric Observatory for Infrared Astronomy (SOFIA) opened the far-infrared view for polarimetry. Using SOFIA/HAWC+, galaxies         \citep{Jones2020, LopezRodriguez2020}, Bok globules \citep{Zielinski2021}, prestellar cores \citep{Redaelli2019} and filaments \citep{Chuss2019} have been observed. \\
In the context of high-mass star formation, highly supercritical filaments are of particular interest. Using the Herschel Space Observatory \citep{Pilbratt2010}, filamentary structures in the interstellar medium have been detected \citep[e.g.,][]{Andre2010, Schisano2020}.
These observational results agree well with the predictions from numerical studies, which are showing that the ISM should be highly filamentary on all scales and star formation is linked to self-gravitating filaments \citep[see][for a review of this topic]{Andre2014}.
Understanding the physical properties of filaments is therefore crucial for understanding star formation in depth on galactic scales.\\
Since the Orion molecular cloud complex is the closest region that is undergoing massive star formation, it has been studied intensively. 
We present polarimetric observations of OMC-3, a star-forming region at a distance of 388\,pc \citep{Kounkel2017}, which is part of the integral shape filament of the Orion molecular cloud. 
Several prestellar and protostellar sources have been identified  in OMC-3 \citep{Chini1997}. The protostellar sources in this region include Class 0 and Class I protostars \citep[e.g.,][]{Chini1997, Nielbock2003}. MMS6 is the brightest source in OMC-3 with at least a factor of five larger flux density at (sub)millimeter wavelengths, as compared to all the other OMC-2/3 sources \citep{Matthews2005b, Takahashi2009}. MMS6  has a bolometric luminosity of $L_\mathrm{bol}$ < 60$\,L_\odot$ and a core mass of $M_\mathrm{core}$ = 30$\,M_\odot$ \citep{Chini1997}. Furthermore, multiple radio jets, molecular outflows, and shock-excited H$_2$ emission have been detected \citep{Yu1997, Reipurth1999, Aso2000, Stanke2002, Matthews2005b}.
In particular, OMC-3 is a well-studied region with polarimetric observations ranging from the far infrared to submillimeter (submm) and millimeter (mm) wavelengths \citep[e.g.,][]{Matthews2001, Houde2004, Takahashi2019, Liu2020}. SCUBA and Hertz polarimetric observations have revealed a highly ordered large-scale magnetic field for OMC-3, perpendicular to the filament.  However, \citet{Takahashi2019} and \citet{Liu2020} showed that the small-scale magnetic field has more complex structures using JVLA and ALMA observations. High-resolution polarimetric observations in the far-infrared enable further insights and restrictions of the properties of the magnetic field and the dust by, for instance, studying the polarization spectrum. The polarization spectrum, namely, the polarization degree as a function of wavelength, was first measured in the far-infrared by \citet{Hildebrand1999} using the Kuiper Airborne Observatory. Since then, a great deal of work has been carried out on the basis of observations \citep[e.g.,][]{Vaillancourt2008, Vaillancourt2012, Gandilo2016} and theoretical frameworks \citep[e.g.,][]{Bethell2007, Draine2009, Guillet2018} in an effort to understand and interpret this quantity. 
Using the SOFIA/HAWC+ polarimeter, it is nowadays possible to study the polarization spectrum more precisely, that is, with higher angular resolution and sensitivity \citep[e.g.,][]{Santos2019, Chuss2019}. 
We report the polarimetric observations of \mbox{OMC-3} at 154\,$\mu$m and 214\,$\mu$m obtained with SOFIA/HAWC+, which provide further insights into the magnetic field properties and the far-infrared polarization spectrum of this  region of interest. 
 
This paper is organized as follows. In Section \ref{Section_Observation}, we describe the data acquisition and reduction and the selection criteria we apply to constrain the data. In Section \ref{Section_Results}, we present the polarization maps of OMC-3 and the corresponding analysis. We derive the magnetic field structure and strength of OMC-3 in Sect. \ref{Section:Magnetic_field}. 
The relation between the polarization degree and the cloud properties is discussed in Sect. \ref{Section:Relationsip_between_polarization_and_cloud_properties}. Additionally, in Section \ref{Section:Magnetic_field_diff_wavelengths}, we provide a short discussion of our findings of the magnetic field structure in the context of complimentary polarimetric observations of this source. Finally, we summarize our results in \mbox{ Section \ref{Section_Conclusion}.}

\section{Observations}  \label{Section_Observation}

\subsection*{Data aquisition} \label{Section:Data_Aquisition}

SOFIA/HAWC+ band D and E observations of OMC-3 were carried out on the 1st of October 2019 as part of the SOFIA Cycle 7 (Proposal 07$\_$0026). Bands D and E provide an angular resolution of 13.6$''$ and 18.2$''$ full width at half maximum (FWHM) at the 154$\,\mu$m and 214$\,\mu$m center wavelength, respectively. The detector format consists of two 64 x 40 arrays, each comprising two 32 x 40 sub-arrays \citep{Harper2019}. The observations were performed using the chop-nod procedure with a chopping frequency of 10.2\, Hz.

The raw data were processed by the HAWC+ instrument team using the data reduction pipeline version 2.3.0. This pipeline consists of different data processing steps including corrections for dead pixels as well as the  intrinsic polarization of the instrument and telescope  \citep[for a brief description of all steps, see for instance][]{Santos2019}, resulting in “Level 4” (science-quality) data. These include FITS images of the total intensity (Stokes $I$),   polarization degree $p$, polarization angle $\theta$, Stokes $Q$ and Stokes $U$, and all related uncertainties. The polarization degree $p$ is calculated via
\begin{equation}
\hspace{3cm} p = \frac{\sqrt{Q^2+ U^2}}{I} \: .
\end{equation}
Furthermore, to increase the reliability of our findings, we apply two additional criteria for the data that will be considered in the subsequent analysis:
\begin{equation}  \label{Formula:Requirement1}
 \hspace{3cm} \frac{I}{\sigma_I}  > 100 \: ,
\end{equation}
\begin{equation} \label{Formula:Requirement2}
\hspace{3cm}  \frac{p}{\sigma_p}  > 3 \: ,
\end{equation}
where $\sigma_I$ and $\sigma_p$ are the standard deviations of $I$ and $p$, respectively. In total, we obtained 1299 and 1710 Nyquist-sampled detections at 154\,$\mu$m and 214\,$\mu$m, respectively, which meet criteria (\ref{Formula:Requirement1}) and (\ref{Formula:Requirement2}).


\section{Results} \label{Section_Results}

\subsection{Polarization map of OMC-3}
\begin{figure*}
   \centering
   \includegraphics[width=\textwidth]{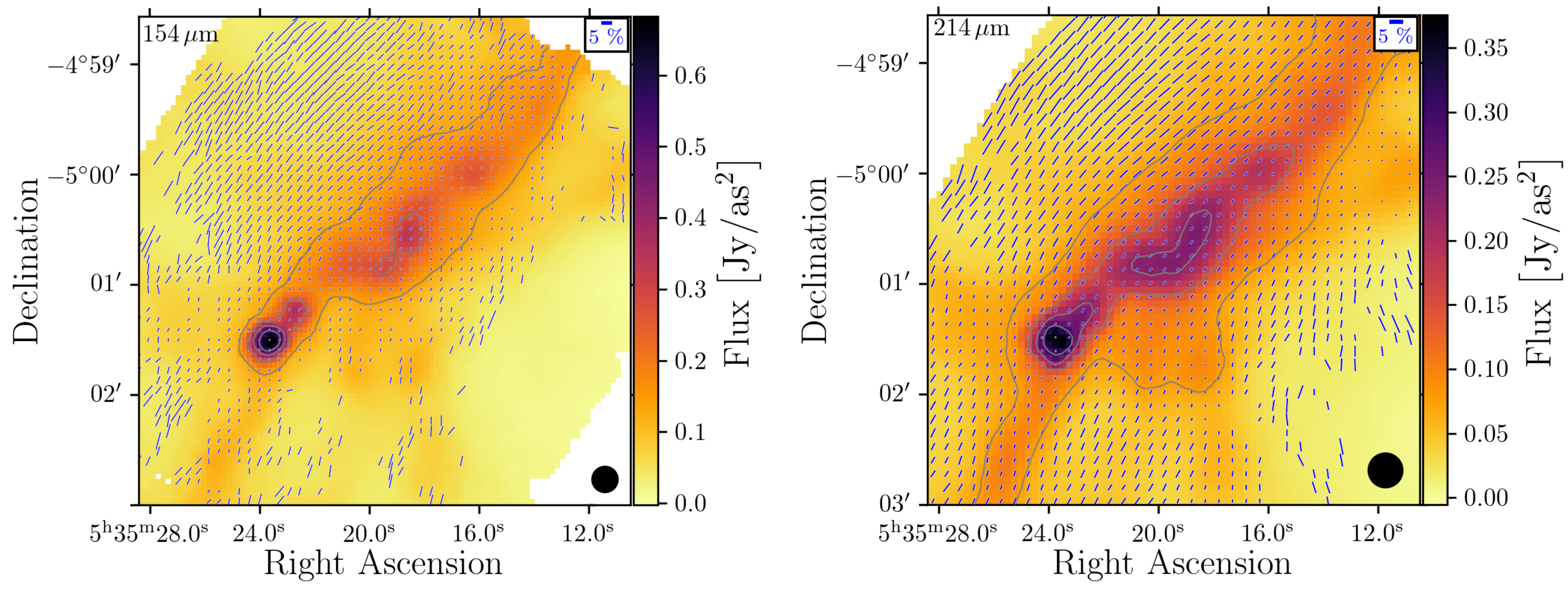}
      \caption{SOFIA/HAWC+ band D (154\,$\mu$m, left) and E (214\,$\mu$m, right) polarization maps of OMC-3. The total intensity is shown with overlaid polarization vectors in blue. The length of the vectors is proportional to the polarization degree and the direction gives the orientation of the linear polarization. The isocontour lines mark 20, 40, 60, and 80$\%$ of the maximum intensity. According to criteria (\ref{Formula:Requirement1}) and (\ref{Formula:Requirement2}), only vectors with $I > 100\, \sigma_I$ and \mbox{$p > 3\, \sigma_p$} ought to be considered (see Sect. \ref{Section:Data_Aquisition}). The beam sizes of 13.6$''$ for 154\,$\mu$m and 18.2$''$ for 214\,$\mu$m (defined by the FWHM) are indicated in the lower right corners of their corresponding plots.
              }
         \label{Polarization_Map_OMC3}
\end{figure*}

The SOFIA/HAWC+ bands D and E have different fields-of-view, namely, 3.7$'$ $\times$ 4.6$'$, and 4.2$'$ $\times$ 6.2$'$ for bands D and E, respectively \citep{Harper2019}. Therefore, we focus on a region with valid polarization data for both bands, that is, polarization data meeting criteria (\ref{Formula:Requirement1}) and (\ref{Formula:Requirement2}). Figure \ref{Polarization_Map_OMC3} shows the resulting band D (154$\,\mu$m, left) and E (214$\,\mu$m, right) polarization maps of OMC-3, overlaid on the corresponding intensity maps. The complete polarization maps for both wavelengths are shown in Fig. \ref{Polarization_Map_OMC3_Complete}.

\begin{figure}
 \includegraphics[width=\hsize]{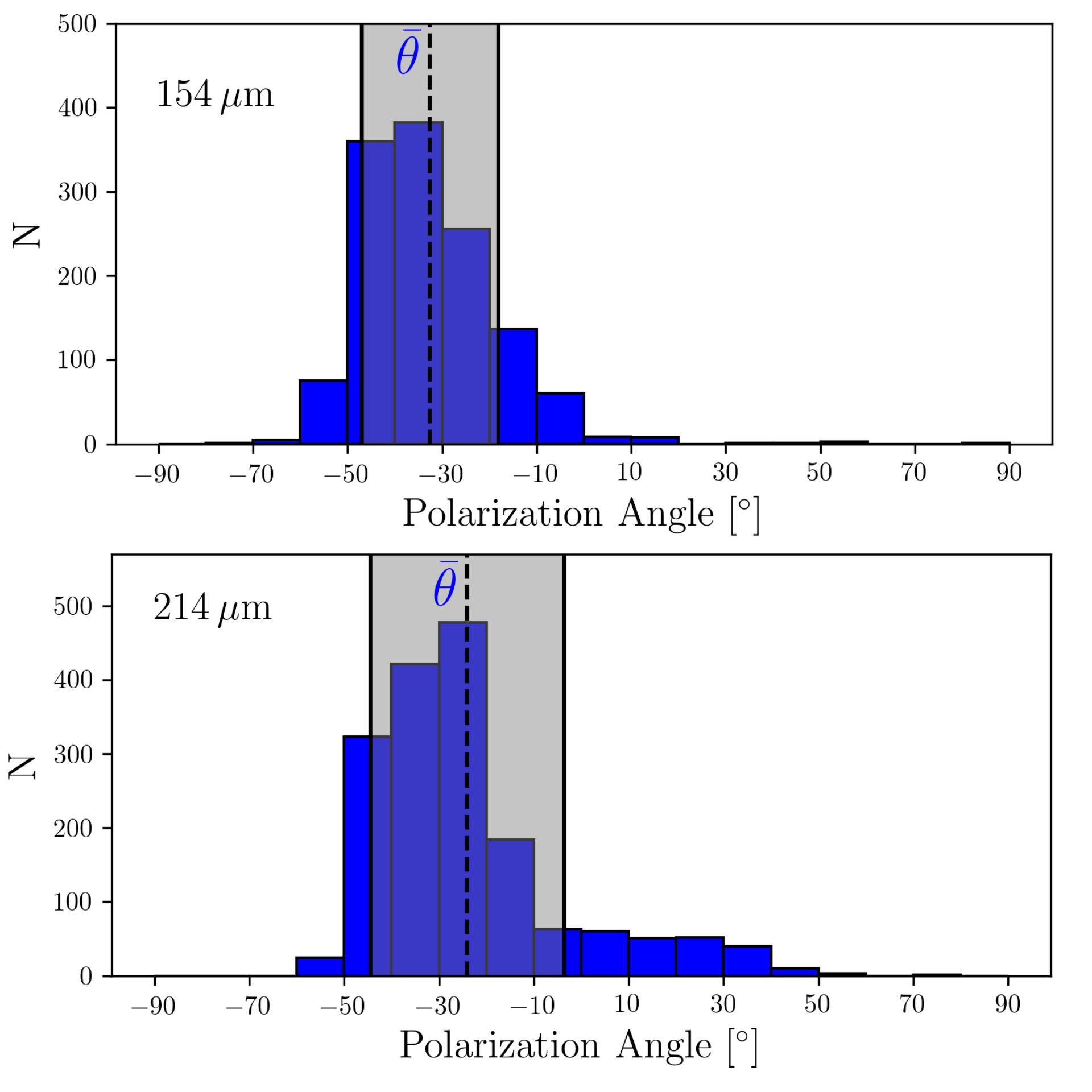}
      \caption{Histograms showing the distribution of polarization angles of band D (154\,$\mu$m, top) and band E (214\,$\mu$m, bottom), respectively. The dashed lines represent the mean polarization angle $\bar{\theta}_{\rm D}$ = -32.6$^\circ$ and \mbox{$\bar{\theta}_{\rm E}$ = -24.1$^\circ$} for 154\,$\mu$m, and 214\,$\mu$m, respectively. The solid lines represent the corresponding 1$\sigma$ levels, 14.5$^\circ$, and 20.4$^\circ$, respectively. 
              }
              \label{Figure:Distribution_Pol_Angles_Pol_Angles}
\end{figure}

Figure \ref{Figure:Distribution_Pol_Angles_Pol_Angles} shows the distribution of the polarization angles $\theta$ of OMC-3 for SOFIA/HAWC+ 154\,$\mu$m (top) and 214\,$\mu$m (bottom). We mostly find polarization angles ranging from -50$^\circ$ to -10$^\circ$ with clear predominance around -30$^\circ$ for both wavelengths. 

\begin{figure}
 \includegraphics[width=\hsize]{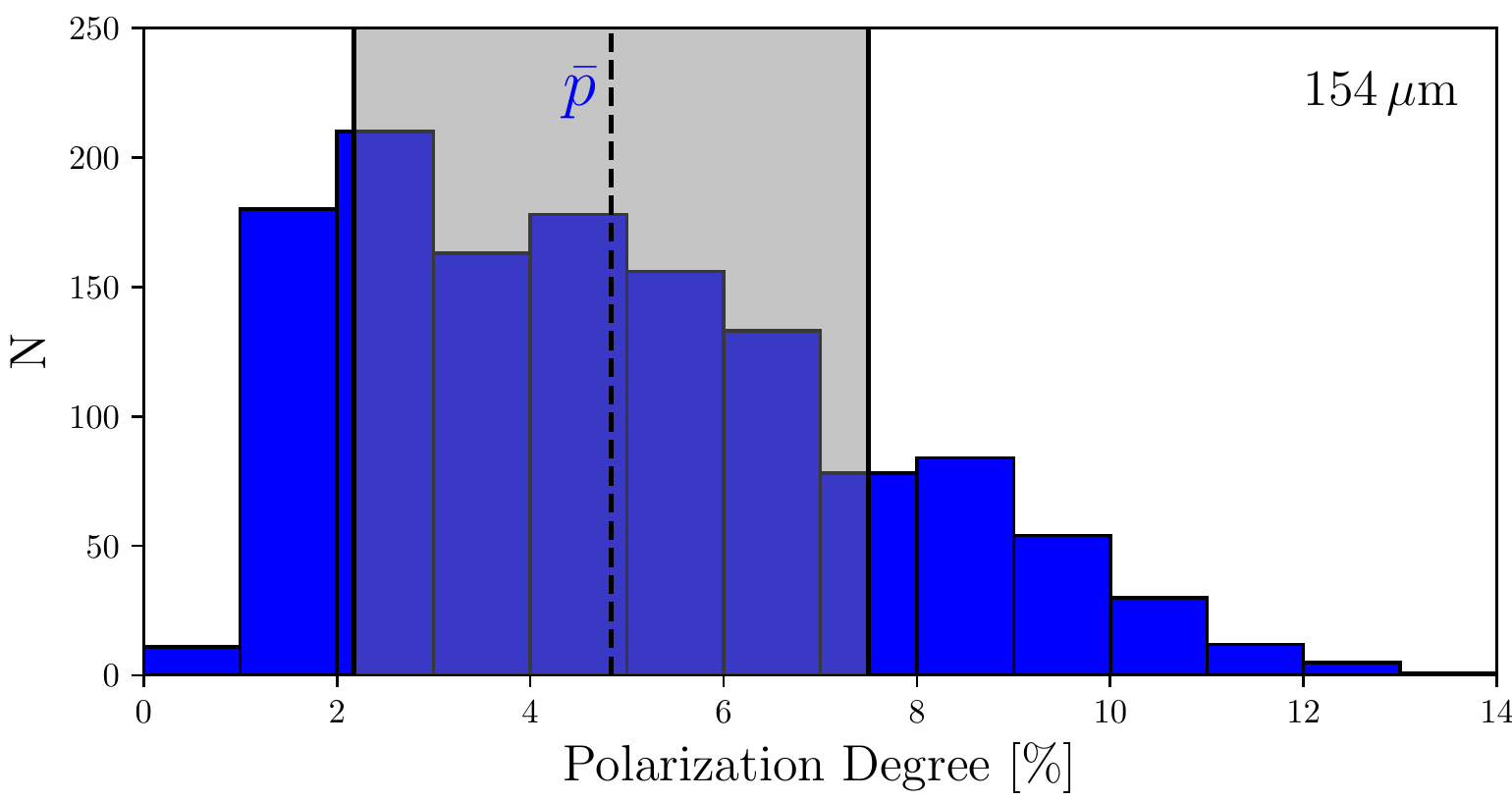}
  \includegraphics[width=\hsize]{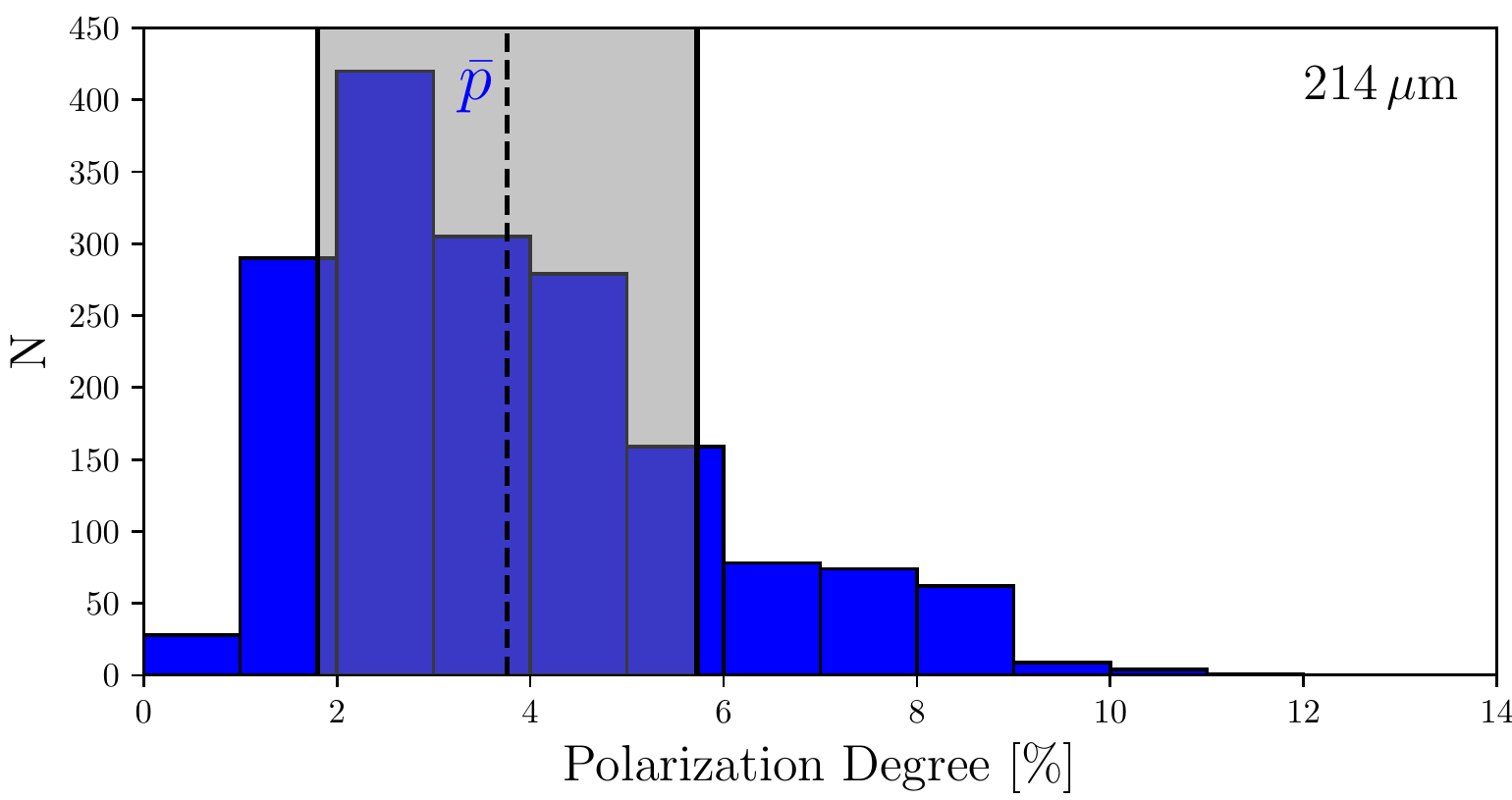}
      \caption{Histograms showing the distribution of polarization degrees of band D (154\,$\mu$m, top) and band E  (214\,$\mu$m, bottom), respectively. The dashed lines represent the mean polarization degree $\bar{p}_{\rm D}$ = 4.8$\,\%$ and \mbox{$\bar{p}_{\rm E}$ = 3.8$\,\%$} for 154\,$\mu$m, and 214\,$\mu$m, respectively. The solid lines represent the corresponding 1$\sigma$ levels, 2.7$\,\%,$ and 2.0$\,\%$, respectively. 
              }
              \label{Figure:Distribution_Pol_Angles_Pol_Degrees}
\end{figure}

In Fig. \ref{Figure:Distribution_Pol_Angles_Pol_Degrees}, the distribution of the polarization degree, $p,$ is shown. For both wavelengths the degree of polarization varies  between 0.5$\,\%$ and $\sim$\,15$\,\%$. The polarization degree is in general higher at 154\,$\mu$m ($\overline{p_\mathrm{154\,\mu m}} = 4.8\,\%$ $\pm$ 2.7$\,\%$) than at 214\,$\mu$m ($\overline{p_\mathrm{214\,\mu m}} = 3.8\,\%$ $\pm$ 2.0$\,\%$ ). 

A higher degree of polarization at a shorter wavelength is not expected, but can be explained by the fact that many polarization vectors at 154\,$\mu$m in regions of higher intensity (lower degree of polarization) do not meet conditions (\ref{Formula:Requirement1}) and (\ref{Formula:Requirement2}). As a result, most of the polarization vectors are in low-intensity (higher degree of polarization) regions. If we only consider regions of higher intensity (e.g., $I > 0.2 \: I_{\mathrm{max}})$, then the degree of polarization is smaller at 154\,$\mu$m than at 214\,$\mu$m \mbox{ ($\overline{p_{\mathrm{154\,\mu m}, \: I\:>\:0.2\:I_\mathrm{max}}} = 2.3\,\%$ $\pm$ 1.0$\,\%$,} \mbox{ $\overline{p_{\mathrm{214\,\mu m}, \: I\:>\:0.2\:I_\mathrm{max}}} = 2.5\,\%$ $\pm$ 1.0$\,\%$).}

\subsection{Magnetic field structure and strength} \label{Section:Magnetic_field}
\begin{figure*}
 \includegraphics[width=\textwidth]{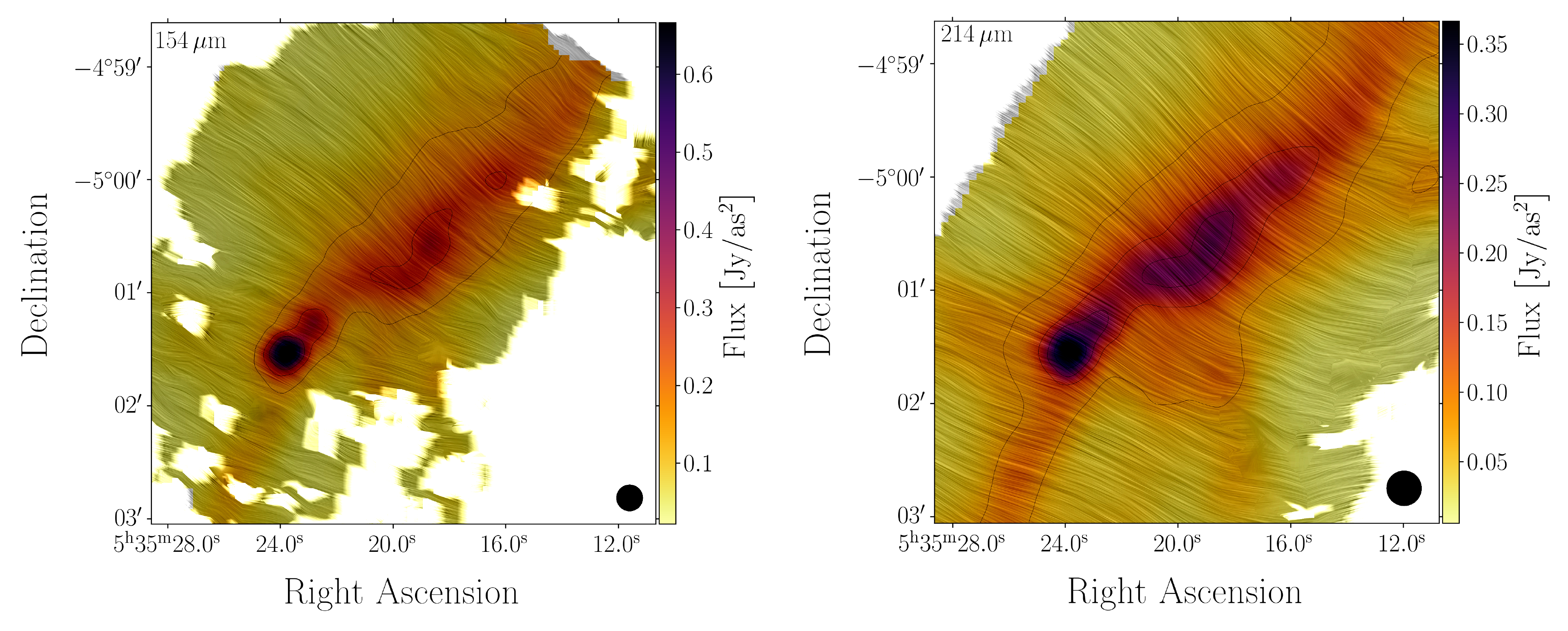}
      \caption{SOFIA/HAWC+ intensity maps of OMC-3 at 154\,$\mu$m (left) and 214\,$\mu$m (right). Using the line-integral-convolution technique, the magnetic field direction is displayed.  The isocontour lines mark 20, 40, 60, and 80$\%$ of the maximum intensity. According to criteria (\ref{Formula:Requirement1}) and (\ref{Formula:Requirement2}), only vectors with $I > 100\, \sigma_I$ and \mbox{$p > 3\, \sigma_p$} are considered (see Sect. \ref{Section:Data_Aquisition}). The beam sizes of 13.6$''$ at 154\,$\mu$m and 18.2$''$ at 214\,$\mu$m (defined by the FWHM) are indicated in the lower right.
              }
              \label{Figure:Magnetic_field}
\end{figure*}

Assuming that the detected polarization is caused by the emission of magnetically aligned non-spherical dust grains, we can rotate the polarization angles by 90$^\circ$ to obtain the projection of the magnetic field direction integrated along the line-of-sight (henceforth, the magnetic field direction).
The magnetic field of OMC-3 is visualized using the line-integral-convolution technique \citep[LIC;][see Fig. \ref{Figure:Magnetic_field}]{Cabral1993}. The intensity is displayed and color-coded, while the LIC textures represent the inferred magnetic field direction. For both wavelengths, the magnetic field direction is oriented perpendicular to the filament structure. This finding is similar to existing polarimetric observations of OMC-3 on similar scales \citep[see Fig. \ref{Figure:OMC3_at_different_wavelengths},][]{Matthews2001, Houde2004}. 

Refering to the Chandrasekhar-Fermi method \citep{ChandrasekharFermi1953}, we  calculate the magnetic field strength of OMC-3 following \citet{Pattle2017}. Here we are meant to assume that the underlying magnetic field is frozen in the cloud material. The plane-of-sky magnetic field strength ($B_\mathrm{pos})$ can be calculated according to \citet{Crutcher2004}:
\begin{equation}
B_\mathrm{pos} = Q \: \sqrt{4 \, \pi \, \rho} \: \frac{\sigma_v}{\sigma_\theta} \approx 9.3 \: \sqrt{n(\mathrm{H}_2)} \: \frac{\Delta v}{\sigma_\theta} \: \mu\mathrm{G}\, ,
\end{equation}
where $\sigma_v$ represents the velocity dispersion, $\sigma_\theta$ the dispersion of the polarization angles, $\rho$ the gas density, $\Delta v$ the velocity dispersion in km s$^{-1}$, $n$(H$_2$) the number density of molecular hydrogen, and $Q$ = 0.5 \citep{Crutcher2004} is a correction factor to account for variation of the field strength on scales smaller than the beam size. 
As the Chandrasekhar-Fermi method does not constrain the line-of-sight component of the magnetic field strength, the total magnetic field strength amounts to:
\begin{equation}
B = \frac{4}{\pi} \: B_\mathrm{pos}.
\end{equation}
For the subsequent analysis, we considered a rectangular area of OMC-3, centered on the point of maximum column density (see Sect \ref{Section:Relationsip_between_polarization_and_cloud_properties}): R.A. 5h35m20.5s, Dec.-5$^\circ$00$'$49$''$.  The selected area has an angular width of 2$'$13.2$''$ and angular height of 1$'$36.2$''$, corresponding to 0.26\,pc and 0.18\,pc, respectively, at a distance of 388\,pc \citep{Kounkel2017}.

\subsubsection{Volume density distribution of OMC-3}
To calculate the magnetic field strength of OMC-3, we determine the volume density distribution, the velocity dispersion, and the dispersion of polarization angles. \\
In Sect. \ref{Section:Relationsip_between_polarization_and_cloud_properties}, we determine the column density across OMC-3 using  a single temperature-modified blackbody fit. By using this fitting technique, which includes, among other things, the optical depth and the Planck function, the column density, the temperature, and the dust emissivity index can be derived.
 To calculate the volume density of OMC-3, we follow \citet{Pattle2017}. We assume that OMC-3 is a cylindrical filament with a radius, $r,$ and length, $L$. The volume then is $\pi r^2 L$ and the above-mentioned rectangular area is the  projection of that volume onto the plane of the sky, were the area is $2rL$.
  The volume density amounts to:
\begin{equation}
n(H_2) = \frac{2\, N(H_2)}{\pi \, r} \: \cos i \, ,
\end{equation}
where $i$ represents the inclination angle between filament and plane of sky. The inclination angle is unknown and we assume that the filament axis is oriented close to the plane of sky, meaning $\cos i \approx$ 1. It should be noted that if the filament is tilted to the plane of sky, the volume density  would decrease by a factor of $\sqrt{2}$, if the inclination angle is 45$^\circ$. Inside the considered area of OMC-3, we obtain a mean value of \mbox{ $N(H_2$) = ($1.71\pm1.0$) $\cdot$ 10$^{22}$ cm$^{-2}$.} We determine the volume density as $n(H_2$) = ($3.82\pm2.24$) $\cdot$ 10$^4$ cm$^{-3}$.

\subsubsection{Velocity dispersion in OMC-3}
\citet{Aso2000} observed the OMC-2/3 region in the H$^{13}$CO$^+$, HCO$^+$ (1-0), and CO (1-0) lines using the Nobeyama 45\,m radio telescope.
We determined the velocity dispersion of the gas in OMC-3 using the  H$^{13}$CO$^+$ observations, as shown in  Table 2 in \citet{Aso2000}. Here, several sources, namely AC2, AC3, AC4, are identified which are located within our considered area.  The mean velocity dispersion amounts to $\sigma_v$ = 0.983 $\pm$ 0.005 km s$^{-1}$.

\subsubsection{Dispersion of polarization angles}
We calculated the standard devation of the mean polarization angles at 154\,$\mu$m and 214\,$\mu$m inside our considered area using the 95$\,\%$ confidence interval. We get $\sigma_{\theta, D}$ = 11.24$^{+0.87}_{-0.75}\:$ deg and \mbox{$\sigma_{\theta, E}$ = 8.68$^{+0.83}_{-0.70}\:$ deg.} In Table \ref{Table_properties_CF_Method}, we provide an overview of the parameters in relation to the calculation of the magnetic field strength of OMC-3. 
Using these values, the corresponding plane-of-sky magnetic field strength amounts to 159\,$\mu$G and 205\,$\mu$G, derived from the 154\,$\mu$m and 214\,$\mu$m measurements, respectively. The total magnetic field strength amounts to 202\,$\mu$G (154\,$\mu$m) and 261\,$\mu$G (214\,$\mu$m). The calculated magnetic field strength values for OMC-3 are lower than the values derived for OMC-1 \citep[300--1000\,$\mu$G,][]{Houde2009, Chuss2019}. The difference can be traced back to the fact that the velocity dispersion for OMC-1 is higher \citep[3.12 km s$^{-1}$, ][]{Pattle2017}. Our derived magnetic field strengths are similiar to 190\,$\mu$G, reported by \citet{Poidevin2010} using 850\,$\mu$m SCUBA data for OMC-3 MMS1-7. \\
Based on the derived magnetic field strength, we calculate the mass-to-flux ratio $\lambda$. We follow \citet{Crutcher2004} to obtain:
\begin{equation}
\lambda = \frac{  (M/\Phi)_\text{observed} }{ (M/\Phi)_\text{crit} } = 7.6 \cdot 10^{-21} \: \frac{N(H_2)}{B} \: \cdot \: \frac{\mu\text{G}}{\text{cm}^2} .
\end{equation} 
Here, $N(H_2)$ describes the column density and $B$ the magnetic field strength. Using the derived values above, we obtain \mbox{ $\lambda_\mathrm{154\,\mu m}$ = 0.64$^{+0.10}_{-0.22}$} and $\lambda_\mathrm{214\,\mu m}$ = 0.49$^{+0.07}_{-0.15}$; in particular,  both measurements indicate that the filament is subcritical, similarly to \mbox{OMC-1}, where \citet{Pattle2017} derived a value of\mbox{ $\lambda_\mathrm{OMC-1}$ = 0.41.}

\begin{table*}
  \begin{center}
    \caption{Overview of the properties in relation to the Chandrasekhar-Fermi method for the calculation of the magnetic field strength of OMC-3.}
    \label{Table_properties_CF_Method}
    \begin{tabular}{ccc}
    \hline \hline    \rule{0pt}{2ex}
      \textbf{Parameter} & \textbf{Symbol} & \textbf{Value}\\
      \hline \hline
      \rule{0pt}{3ex}
       \rule{0pt}{2ex}
      \noindent Hydrogen column density  & $N(H_2)$ & $(1.70\pm1.0)$ $\cdot$ 10$^{22}$ cm$^{-2}$\\
      \rule{0pt}{2ex}
      Hydrogen volume density  & $n(H_2)$ & $(3.81\pm2.24)$ $\cdot$ 10$^{4}$ cm$^{-3}$\\
       \rule{0pt}{2ex}\rule{0pt}{3ex}
      Angular dispersion (154\,$\mu$m) &  $\sigma_{\theta, 154\,\mu\mathrm{m}}$ & 11.24$^{+0.87}_{-0.75}\:$ deg\\

       \rule{0pt}{2ex} \rule{0pt}{3ex}
      Angular dispersion (214\,$\mu$m) &  $\sigma_{\theta, 214\,\mu\mathrm{m}}$ & 8.68$^{+0.83}_{-0.70}\:$ deg\\
       \rule{0pt}{2ex}\rule{0pt}{3ex}
      Velocity  dispersion\,$^a$  & $\Delta v$ & 0.98 km s$^{-1}$\\

       \rule{0pt}{2ex}\rule{0pt}{3ex}
      POS magnetic field strength (154\,$\mu$m) & $B_{\mathrm{pos, 154\,\mu\mathrm{m}}}$  & 158.6$^{+58.9}_{-63.3}\:\mu$G \\
      \rule{0pt}{2ex}\rule{0pt}{3ex}
      POS magnetic field strength (214\,$\mu$m) & $B_{\mathrm{pos, 214\,\mu\mathrm{m}}}$  & 205.4$^{+82.0}_{-83.6}\:\mu$G \\
      \rule{0pt}{2ex}\rule{0pt}{3ex}
      Total magnetic field strength (154\,$\mu$m) & $B_{\mathrm{154\,\mu\mathrm{m}}}$  & 201.9$^{+75.0}_{-80.6}\:\mu$G  \\
      \rule{0pt}{2ex}\rule{0pt}{3ex}
      Total magnetic field strength (214\,$\mu$m) & $B_{\mathrm{214\,\mu\mathrm{m}}}$  & 261.4$^{+104.4}_{-106.4}\:\mu$G\\ 
      \hline \hline
    \end{tabular}
     \tablefoot{\footnotesize $^a$\citealt{Aso2000}}
  \end{center}
\end{table*}


\subsection{Correlation between magnetic field structures and cloud properties} \label{Section:Relationsip_between_polarization_and_cloud_properties}

In the following, we investigate how the polarization degree and angle, measured with SOFIA/HAWC+, change with cloud properties, namely, the column density and temperature. For this purpose, we follow \citet{Chuss2019} to construct column density, temperature, and dust emissivity maps.  

\subsubsection*{Data preparation} \label{Subsection:Data_preparation}
In order to derive the column density, temperature, and dust emissivity maps, we used the SOFIA/HAWC+ 154\,$\mu$m and 214\,$\mu$m data, together with JCMT/SCUBA-2\footnote{Observation ID scuba2$\_$00015$\_$20180906T152402} 850\,$\mu$m and Herschel PACS\footnote{Observation IDs 1342204250, 1342204251} 70$\,\mu$m and 160$\,\mu$m data. We re-projected all data to the pixel scale of the measurement of 214\,$\mu$m. In the next step, we beam-convolved the 70, 154, 160, and 850\,$\mu$m data to the corresponding resolution of HAWC+ band E (214\,$\mu$m) of 18.2$''$.

\subsubsection*{Fitting routine}
\begin{figure*}
 \includegraphics[width=\textwidth]{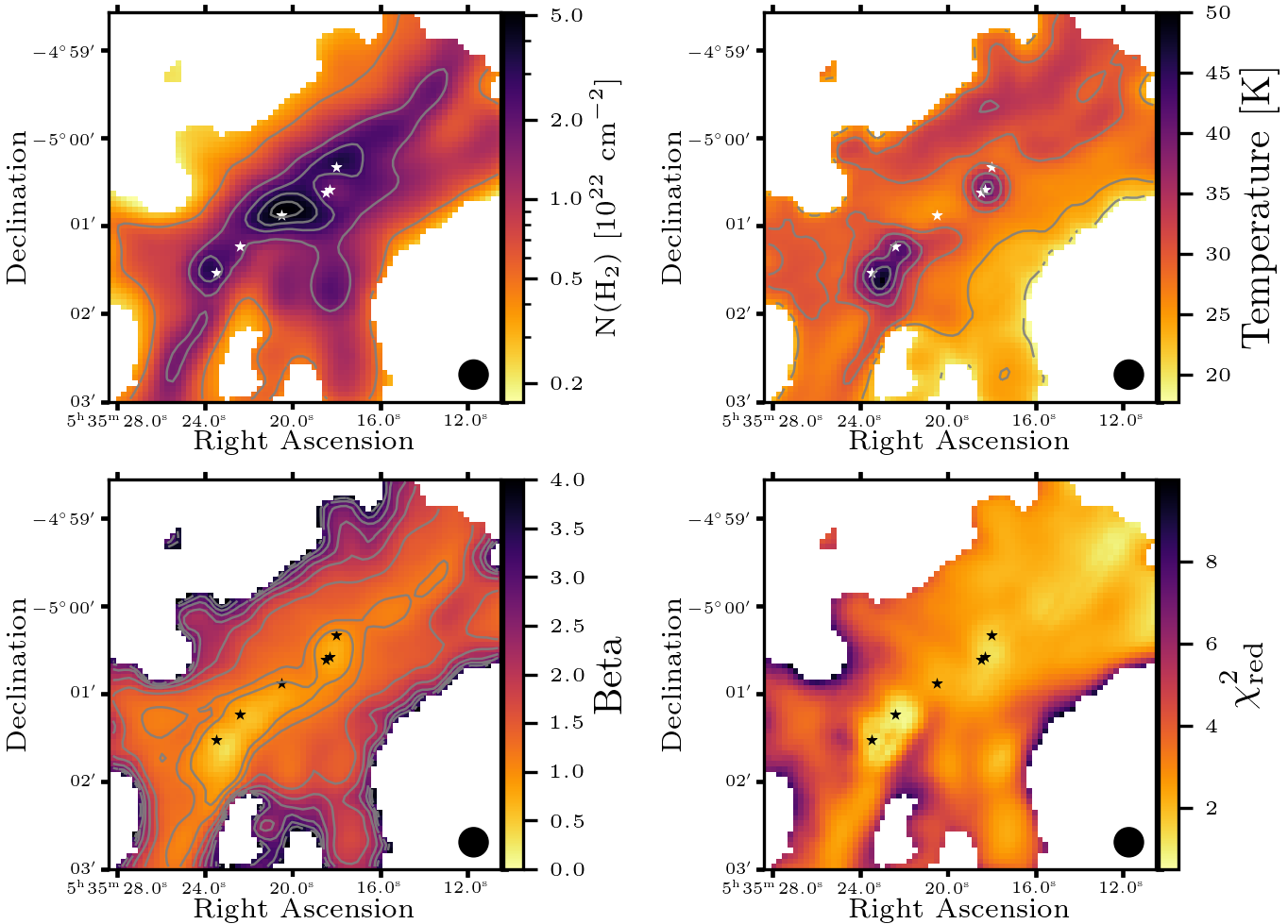}
      \caption{Maps of column density (top left), temperature (top right), dust emissivity index (bottom left), and the corresponding reduced $\chi^2$ (bottom right). The beam size of 18.2$''$  (band E, 214\,$\mu$m) is indicated in the lower right for each figure.
              }
              \label{Figure:Column_density_map}
\end{figure*}

We fit a single-temperature modified blackbody function to each pixel:
\begin{equation}
I_\nu = ( 1 - \exp (-\tau_\nu)) \: B_\nu (T)\, .
\end{equation}
Here, $B_\nu (T)$ describes the Planck function and $\tau_\nu$ the optical depth:
\begin{equation}
\tau_\nu = \epsilon \: (\nu/\nu_0)^\beta \, ,
\end{equation}
where the quantity $\beta$ is the dust emissivity index and $\nu_0$ a reference frequency. The parameter $\epsilon$ is a scaling factor related to the column density $N(H_2)$:
\begin{equation}
\epsilon = \kappa_{\nu_0} \: \mu \: m_H \: N(H_2) \, .
\end{equation}
Here, $\kappa_{\nu_0}$ is the reference dust opacity per unit mass, $m_H$ the atomic mass of hydrogen, and $\mu$ the mean molecular weight per hydrogen atom.
We set $\nu_0$ to 1000\,Hz and adopt \mbox{$\kappa_{\nu_0}$ (1000 Hz) = 0.1 cm$^2$ g$^{-1}$.} The final resulting fit function is:
\begin{equation}
I_\nu = \left( 1 - \exp \left(- \kappa_{\nu_0} \: \mu \: m_H \: N(H_2) \: \left(\nu/\nu_0\right)^\beta \right)  \right) \: \frac{2h \nu^3}{c^2} \: \frac{1}{\exp\left(\frac{h \nu}{k T}\right) -1} \, .
\end{equation}
The fit parameters are the column density $N(H_2)$, the temperature $T,$ and the dust emissivity index $\beta$. We adopted the uncertainties for the Stokes I data given in \citet{Chuss2019} and rejected all fitted pixels with a reduced $\chi^2$ (used to estimate the goodness of the fit) greater than 10. 
 The resulting maps of column density, temperature, and dust emissivity index are shown in Fig. \ref{Figure:Column_density_map}.

The dust emissivity index $\beta$ is lowest at the central regions of OMC-3 and increases toward the outer regions, indicating potential dust grain growth at regions with higher density and in the vicinity of stellar sources. The mean dust emissivity index amounts to \mbox{$\bar{\beta}$ = 1.72.} With this type of fitting technique, it is often omitted that beta is a free parameter \citep[e.g., ][]{Gandilo2016, Santos2019}. \citet{Santos2019} used a fixed value of $\beta$ = 1.62 for $\rho$ Oph A, while \citet{Gandilo2016} applied $\beta$ = 2.0 for the Vela C molecular cloud. These fixed values are similar to our derived mean value. \\
In the maps shown in Fig. \ref{Figure:Column_density_map}, we mark the known stellar sources \citep[white/black star signs; from][]{Chini1997}.
 The embedded stellar sources radiate and heat the surrounding dust, that is, the position of stellar sources is closely connected to an increased temperature. The highest temperature can be found in the vicinity of MMS6, the most luminous source in OMC-3 \citep{Matthews2005b, Takahashi2009}. The only exception where the presence of a stellar source is not related to an increased temperature, is MMS4 at R.A.: 5h 35m 20.5s, Dec.: -5$^\circ$0$'$53$''$.  One possible explanation for this sole exception is the optical depth, which is highest at this stellar position (see Fig. \ref{Figure:Maps_Optical_Depth} in the Appendix). The mean temperature is $\overline{T}$ = 28.38\,K. The column density is highest around MMS4 with \mbox{$N(H_{2})_{\mathrm{max}}$ = 5.15 $\cdot$ 10$^{22}$ cm$^{-2}$.} The mean column density amounts to $\overline{ N(H_{2}) }$ = 1.10 $\cdot$ 10$^{22}$ cm$^{-2}$.

\subsubsection{Degree of polarization as a function of the local column density}
\begin{figure}
\includegraphics[width=\hsize]{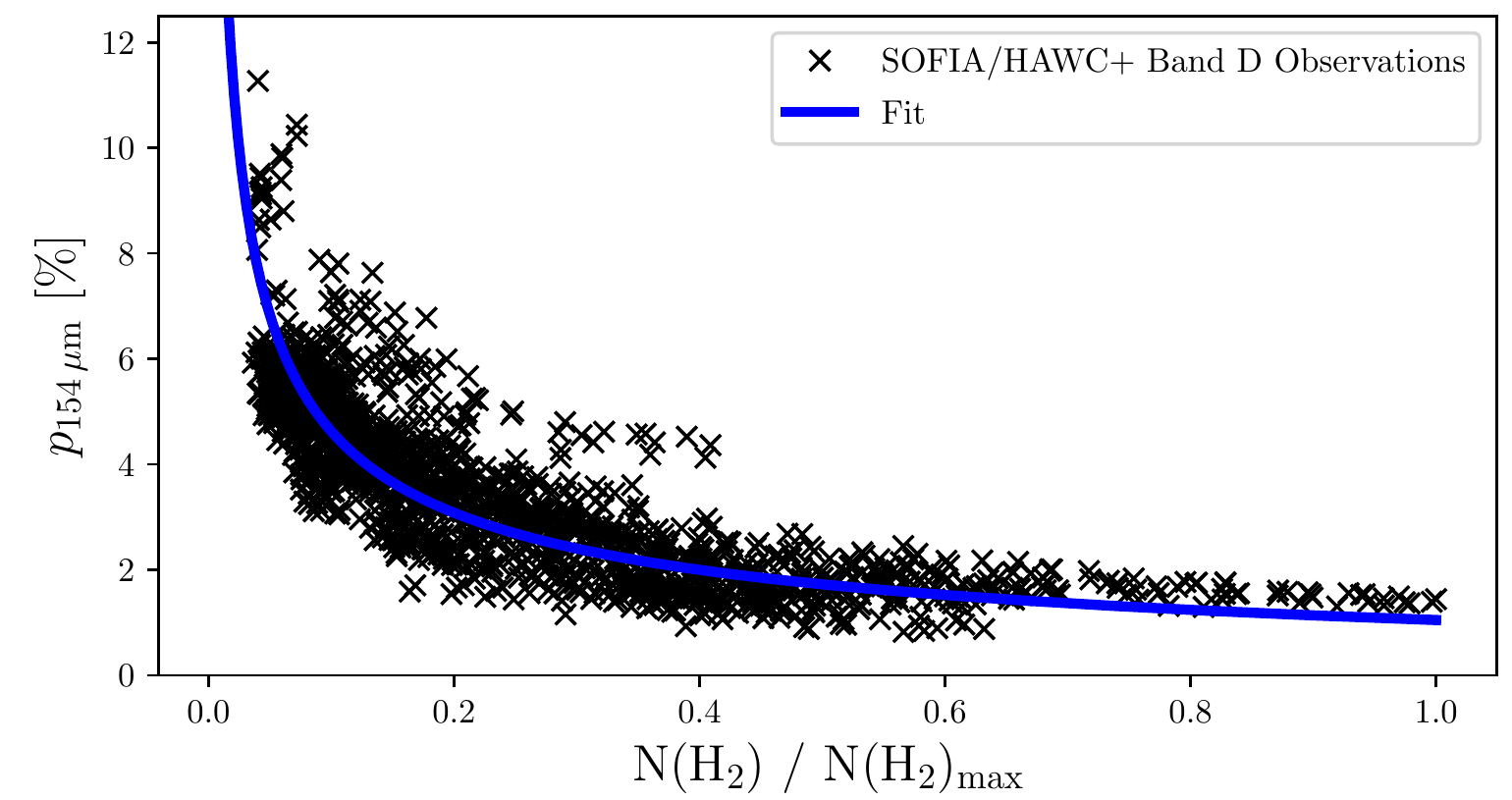}
\includegraphics[width=\hsize]{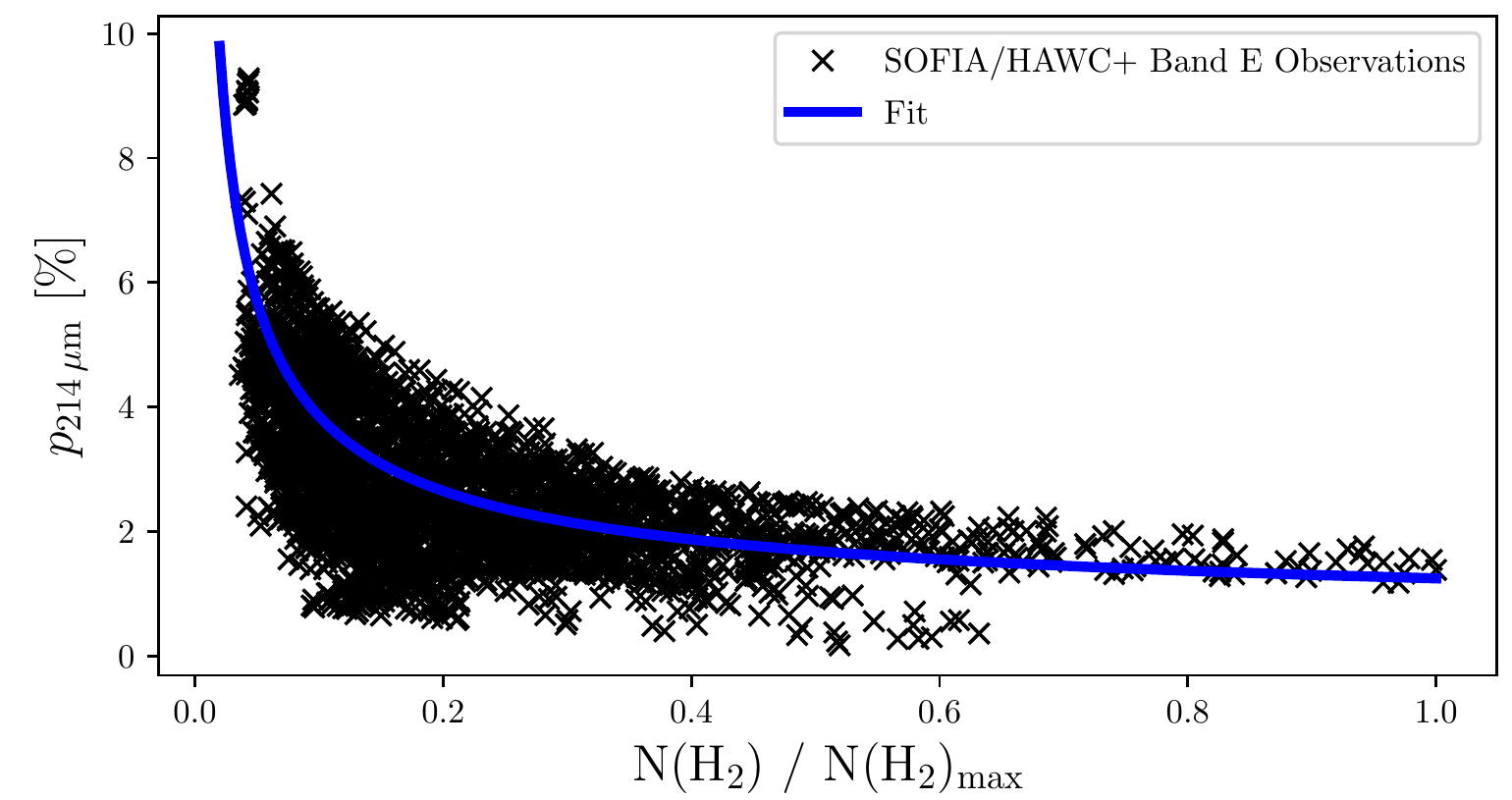}
\caption{Polarization degree at 154\,$\mu$m (top) and 214\,$\mu$m (bottom) as a function of the column density, scaled to the maximum value.} \label{Figure:Pol_Degree_vs_Column_Density}
\end{figure}

Polarimetric observations of star-forming regions often show a decreasing degree of polarization with increasing density, making up so-called "polarization holes" \citep[e.g.,][]{Henning2001, Wolf2003, Chuss2019, Zielinski2021}. We investigate how the polarization degree changes in relation to the column density derived above.
As in the data preparation shown in  Sect. \ref{Subsection:Data_preparation}, the Stokes parameter $Q$ and $U$ at 154\,$\mu$m are re-projected and beam-convolved to 214\,$\mu$m resolution as well. The polarization degree is then calculated using
\begin{equation}
p = \frac{\sqrt{Q^2+ U^2}}{I} \: .
\end{equation}
Our results for 154\,$\mu$m and 214\,$\mu$m are shown in Fig. \ref{Figure:Pol_Degree_vs_Column_Density}.

For both wavelengths, the polarization degree decreases from $\sim$ 10$\%$ to $\lesssim$ 1$\%$ with increasing density. Inspired by the work of \citet{Davis2000}, who applied linear least-square fits to polarimetric observations of the Serpens cloud core and found a correlation between the measured polarization and intensity, \citet{Henning2001} found a correlation between these quantities in the case of the two Bok globules CB54 and DC253-1.6 as well. 
They approximated the decrease in the polarization degree as a function of increasing intensity using the following equation:
\begin{equation}
p = a_0 + a_1 \cdot \left( \frac{I}{I_\text{max}} \right)^{a_2}, 
\end{equation}
where  $a_0 , a_1$, and $a_2$ are fitting parameters. We applied the same technique, but refer to the column density instead of intensity. If we assume that the dust distribution is optically thin, the intensity is connected to the density. Therefore, the subsequent results are comparable. For our case we apply
\begin{equation}
p = a_0 + a_1 \cdot \left( \frac{ N(H_2)  }{N(H_2)_\text{max}} \right)^{a_2}.
\end{equation}
We obtained $a_{0, 154\,\mu\mathrm{m}}$ = -0.53 $\pm$ 0.24 , \mbox{$a_{1, 154\,\mu\mathrm{m}}$ = 1.58 $\pm$ 0.20,} $a_{2, 154\,\mu\mathrm{m}}$ = -0.51 $\pm$ 0.04 and $a_{0, 214\,\mu\mathrm{m}}$ = 0.44 $\pm$ 0.16 , \mbox{$a_{1, 214\,\mu\mathrm{m}}$ = 0.81 $\pm$ 0.12, } $a_{2, 214\,\mu\mathrm{m}}$ = -0.63 $\pm $ 0.05. The parameter $a_2$, which describes the slope of the curve, is slightly higher at 214\,$\mu$m. Interestingly, comparing this slope to those reported in the previous studies for Bok globules, a different object class, one can see that the slopes are similar. For the Bok globule B335, \citet{Wolf2003} derived \mbox{$a_{2, 850\,\mu\mathrm{m}}$ = -0.43} with JCMT/SCUBA and \citet{Zielinski2021} derived \mbox{$a_{2, 214\,\mu\mathrm{m}}$ = -0.55} with SOFIA/HAWC+. Furthermore, the calculated slope for the Bok globule CB54 is $a_{2, 850\,\mu\mathrm{m}}$ = -0.64 \citep{Henning2001} and $a_{2, 850\,\mu\mathrm{m}}$ = -0.55 for DC 253-1.6 \citep{Henning2001}. See Table \ref{Table_values_polarization_hole} for an overview of all calculated $a_2$ values. 
There are multiple possible reasons for the occurence of polarization holes. Since the slope is similar for different object classes and wavelengths, this may be a hint that the same effects are responsible. However, this needs further investigation, since the spatial scale on which the polarization hole in OMC-3 is examined is larger than that of the Bok Globule studies mentioned above.

\begin{table*}
  \begin{center}
    \caption{Calculated values for the parameter $a_2$ that describes the slope of the polarization hole at different wavelengths.}
    \label{Table_values_polarization_hole}
    \begin{tabular}{ccccc}
    \hline \hline    \rule{0pt}{2ex}
      \textbf{Object} & \textbf{Wavelength} & \textbf{Instrument} & $\mathbf{a_2}$ &  \textbf{Reference}\\
      \hline \hline
      \rule{0pt}{3ex}
      \noindent OMC-3  &  154\,$\mu$m & SOFIA/HAWC+  & -0.51 & this paper \\
      \rule{0pt}{2ex}
      OMC-3  &  214\,$\mu$m  & SOFIA/HAWC+ & -0.63 &  this paper\\
      \rule{0pt}{2ex}
      B335  &  214\,$\mu$m &  SOFIA/HAWC+  &  -0.55 &  \citet{Zielinski2021} \\
      \rule{0pt}{2ex}
      B335  &  850\,$\mu$m &  JCMT/SCUBA & -0.43 &  \citet{Wolf2003}  \\
      \rule{0pt}{2ex}
      CB54  &  850\,$\mu$m & JCMT/SCUBA & -0.64 & \citet{Henning2001} \\
      \rule{0pt}{2ex}
      DC 253-1.6  &  850\,$\mu$m & JCMT/SCUBA & -0.55 &  \citet{Henning2001} \\

      \hline \hline
    \end{tabular}
  \end{center}
\end{table*}

\subsubsection{Polarization hole in OMC-3}
\begin{figure}
\includegraphics[width=\hsize]{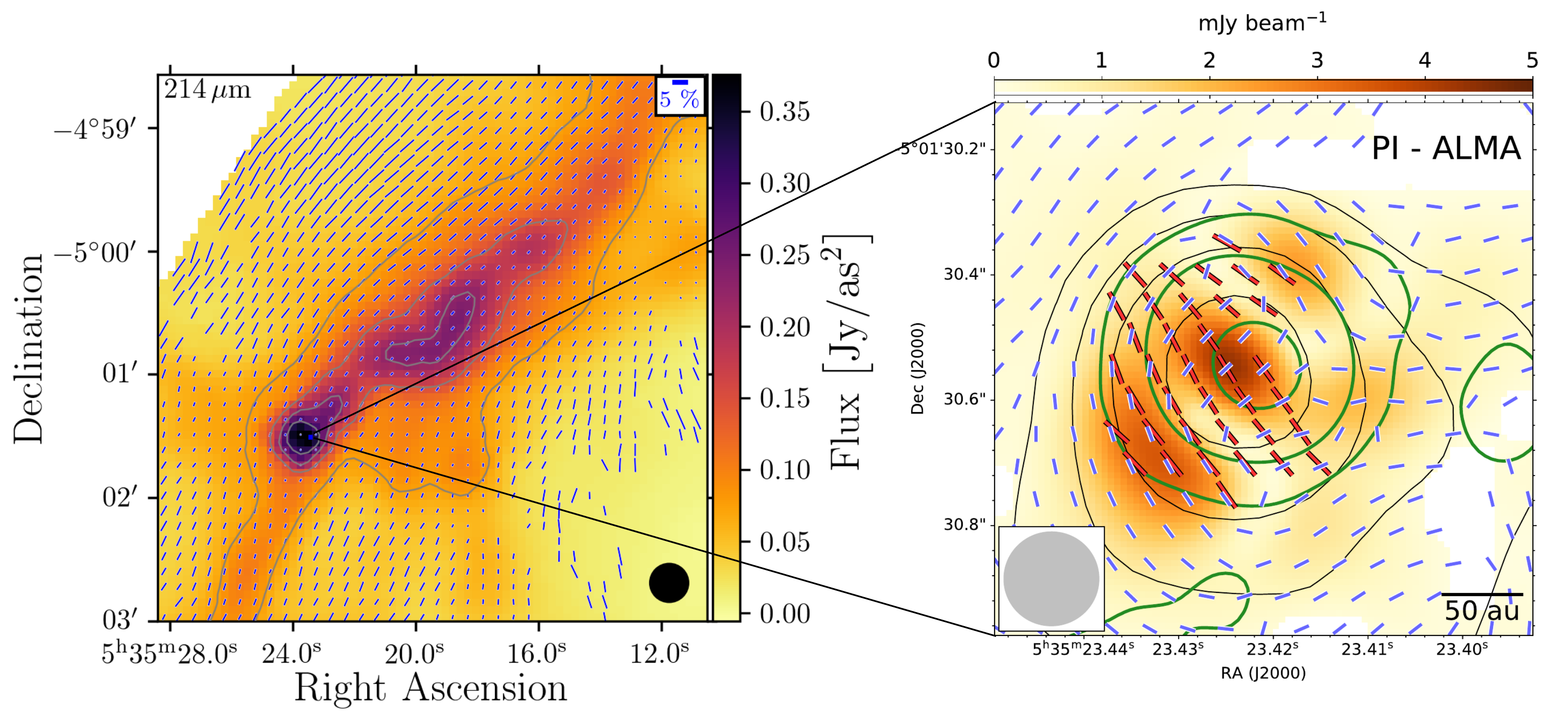}
\caption{OMC-3 polarization maps at different scales. Left: The total intensity is shown with overlaid polarization vectors in blue. The length of the vectors is proportional to the polarization degree and the direction gives the orientation of the linear polarization. Right: Total intensity, observed at 1.2\,mm with ALMA, is shown with overlaid polarization vectors in blue. Red polarization vectors are oberserved with JVLA at 9\,mm \citep[][\copyright AAS. Reproduced with permission]{Liu2020}.} \label{Figure:OMC3_Band_E_ALMA}
\end{figure}

The SOFIA/HAWC+ observations show a decrease of the polarization degree toward dense regions of OMC-3. As mentioned above, there are several existing hypotheses that are aimed at explaining this phenomenon, such as an insufficient angular resolution of a possibly complex magnetic field structure on scales below the resolution of the polarization maps \citep[e.g.,][]{Shu1987,Wolf2004}, a disruption of spinning larger grains into smaller fragments \citep[radiative torque disruption, ][]{Hoang2019A, Hoang2019B}, or certain combinations of optical depth,  dust grain size, and chemical composition \citep{BrauerWolfReissl}.  What is particulaly interesting for our purposes is the latter proposal. \citet{BrauerWolfReissl} showed that a polarization hole can occur as a result of the superposition of polarized emission and dichroic extinction, which act as counteracting mechanisms. This effect may even cause a flip in the polarization direction by 90$^\circ$, if the dichroic extinction dominates over dichroic emission \citep{Reissl2014, BrauerWolfReissl}. Indeed, \citet{Liu2020} showed this 90$^\circ$ flip for OMC-3 (MMS6) with a comparison of 1.2\,mm ALMA and 9\,mm JVLA polarimetric observations, see Fig. \ref{Figure:OMC3_Band_E_ALMA}. They find that the innermost $\sim$ 100\,au region of OMC-3 (MMS6) is optically thick at 1.2\,mm and optically thin at 9\,mm.

Given a moderate optical depth, the counteracting mechanisms of polarized emission and absorption result in a decrease of the polarization degree.  However, this effect can only be applied to explain the polarization hole in the innermost area of OMC-3. The reason for the decrease in the degree of polarization in the outer areas is unknown. The polarimetric ALMA observation shows that the magnetic field in OMC-3 is more complex on smaller scales. Due to beam-averaging, an unresolved and more complex magnetic field would lead to a lower degree of polarization. While the optical depth seems to have an effect on the decrease in the polarization degree, we cannot rule out a potential influence of the magnetic field complexity.  However, other effects, such as the radiative torque disruption  \citep[][]{Hoang2019A, Hoang2019B} or less-aligned dust grains at higher densities \citep{Goodman1992, Creese1995},  along with  their contribution to the polarization hole, cannot be ruled out.

\subsubsection{Degree of polarization versus temperature}
\begin{figure}
\includegraphics[width=\hsize]{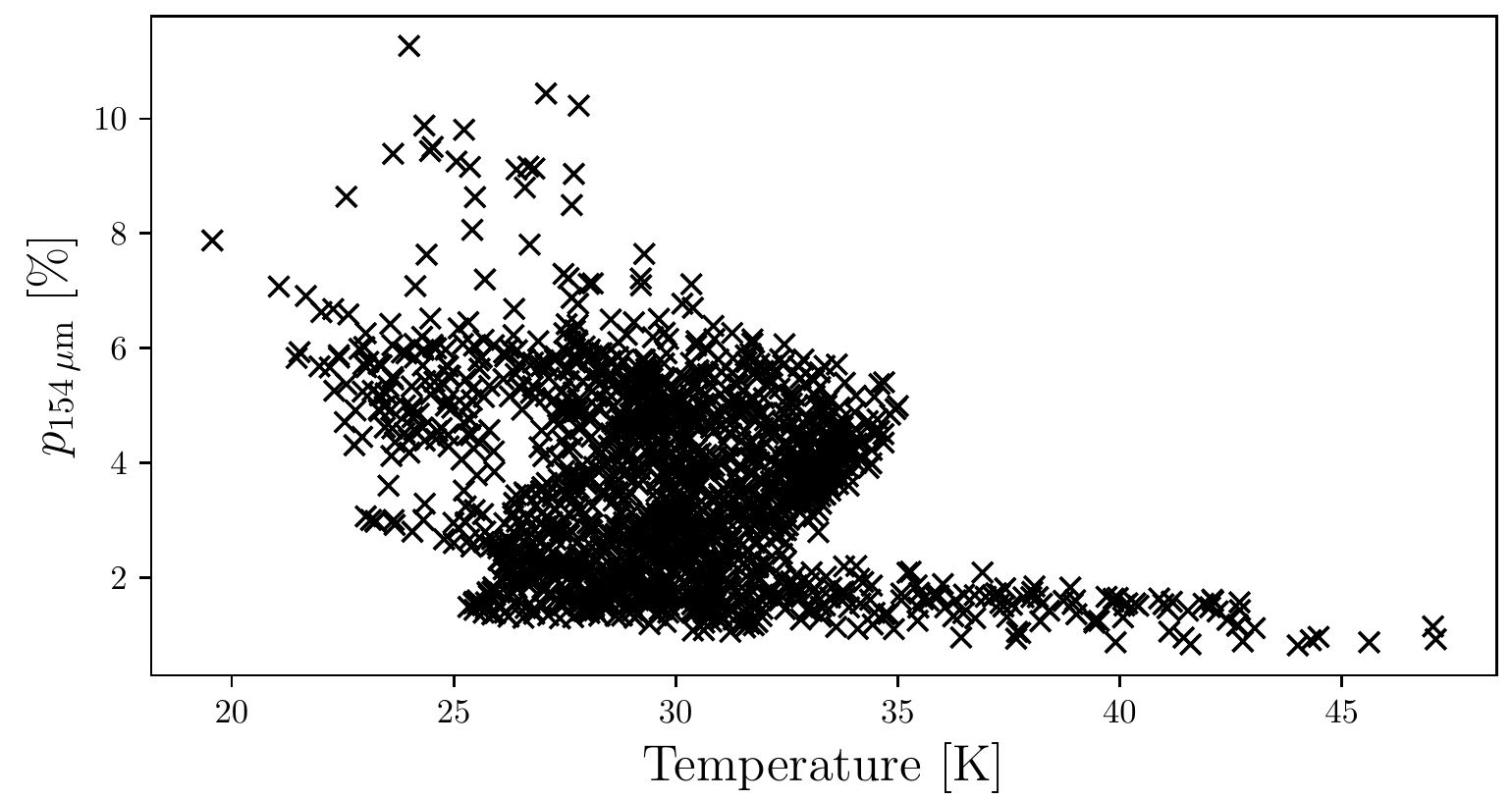}
\includegraphics[width=\hsize]{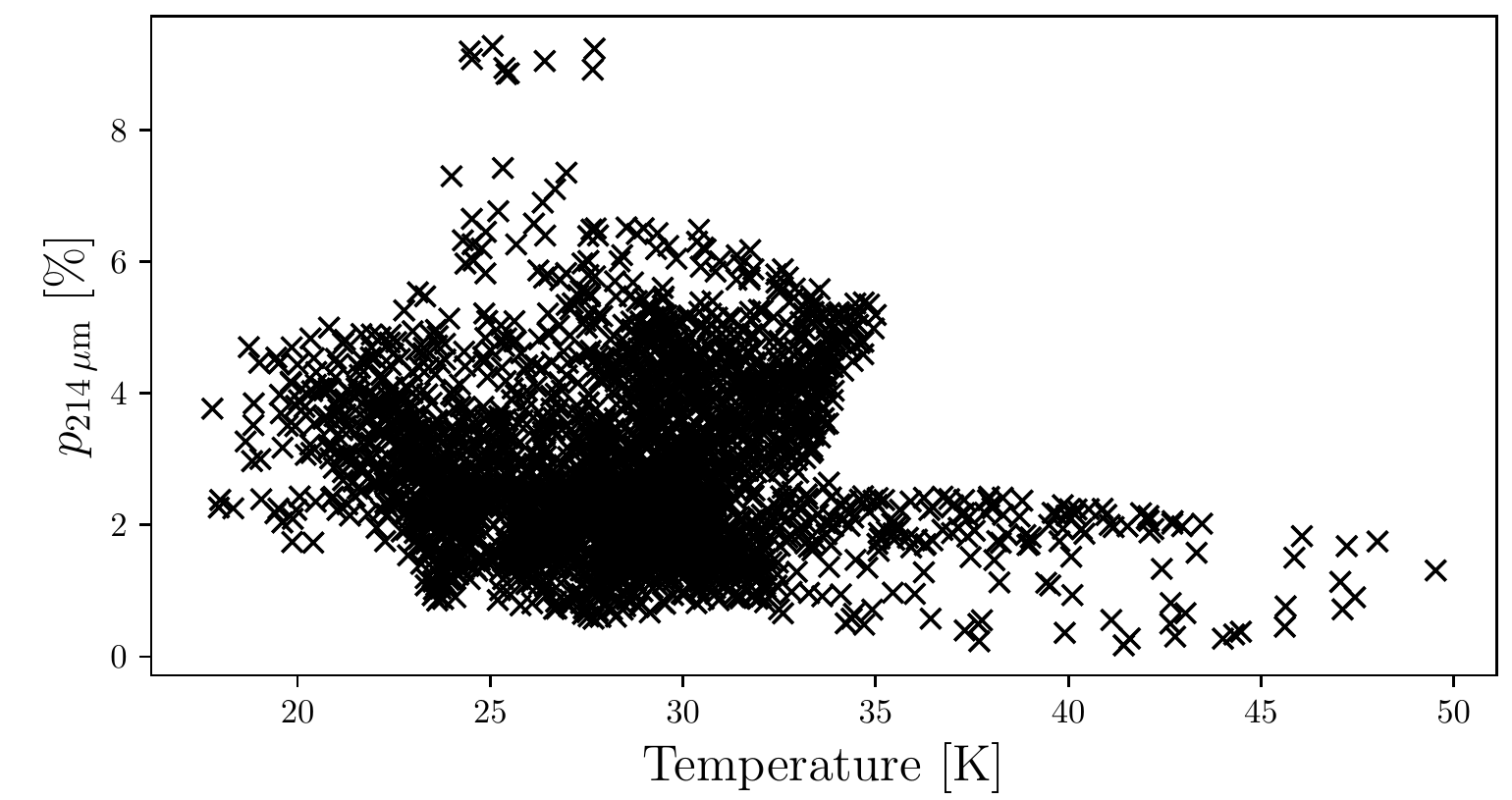}
\caption{Polarization degree at 154\,$\mu$m (top) and 214\,$\mu$m (bottom) as a function of the temperature.} \label{Figure:Pol_Degree_vs_Temperature}
\end{figure}

In Fig. \ref{Figure:Pol_Degree_vs_Temperature}, we show the obtained polarization degrees at 154\,$\mu$m and 214\,$\mu$m as  a function of the derived temperature. The temperature is strongly correlated with the position of stars. The degree of polarization drops sharply in the vicinity of the stars, while the degree of polarization is higher in colder regions.

\subsubsection{Polarization spectrum versus column density and temperature}
In the next step, we investigate how the polarization spectrum, that is, the polarization degree as a function of wavelength, changes with column density and temperature of OMC-3. We define the polarization spectrum as $p_{\mathrm{214\,\mu m}} / p_{\mathrm{154\,\mu m}}$, similiar to \citet{Santos2019}, who studied the polarization spectrum of \mbox{$\rho$ Oph A} and \citet{Michail2021}, who studied the polarization spectrum of OMC-1.  Using this definition, $p_{\mathrm{214\,\mu m}} / p_{\mathrm{154\,\mu m}}$ < 1 indicates a negative spectral slope and $p_{\mathrm{214\,\mu m}} / p_{\mathrm{154\,\mu m}}$ > 1 a positive spectral slope. We calculated the polarization spectrum for all pixels where we have valid data for the polarization degrees at both wavelengths, namely, those data meeting criteria \ref{Formula:Requirement1} and \ref{Formula:Requirement2}, and valid data for the column density and temperature.

\begin{figure}
\includegraphics[width=\hsize]{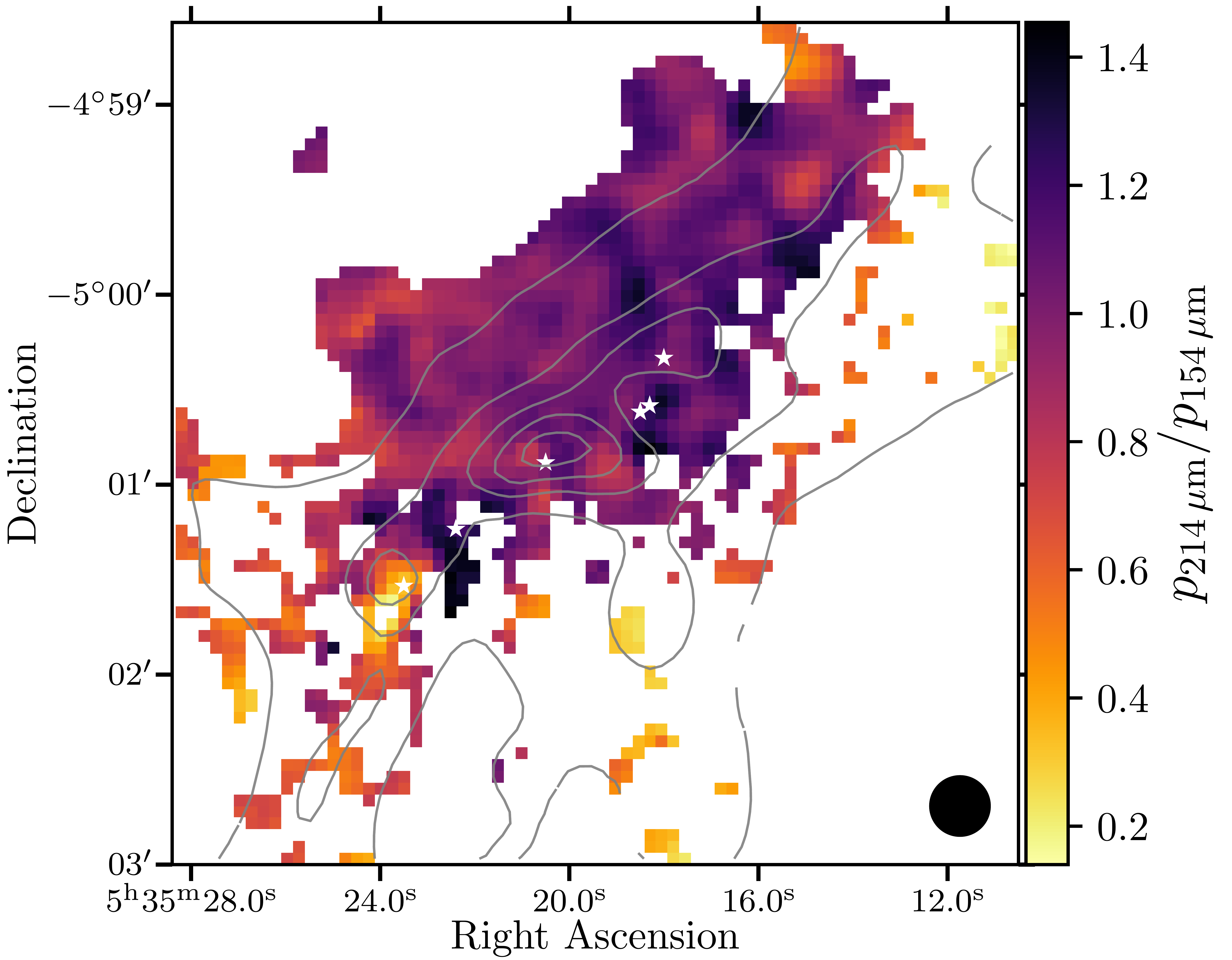}
\caption{Polarization spectrum ($p_{\mathrm{214\,\mu m}} / p_{\mathrm{154\,\mu m}})$ map of OMC-3. The isocontour lines mark 10, 30, 50, 70, and 90$\%$ of the maximum column density. The beam size of  18.2$''$ at 214\,$\mu$m (defined by the FWHM) is indicated in the lower right.} 
\label{Figure:Map_polarization_spectrum}
\end{figure}

\begin{figure}
\includegraphics[width=\hsize]{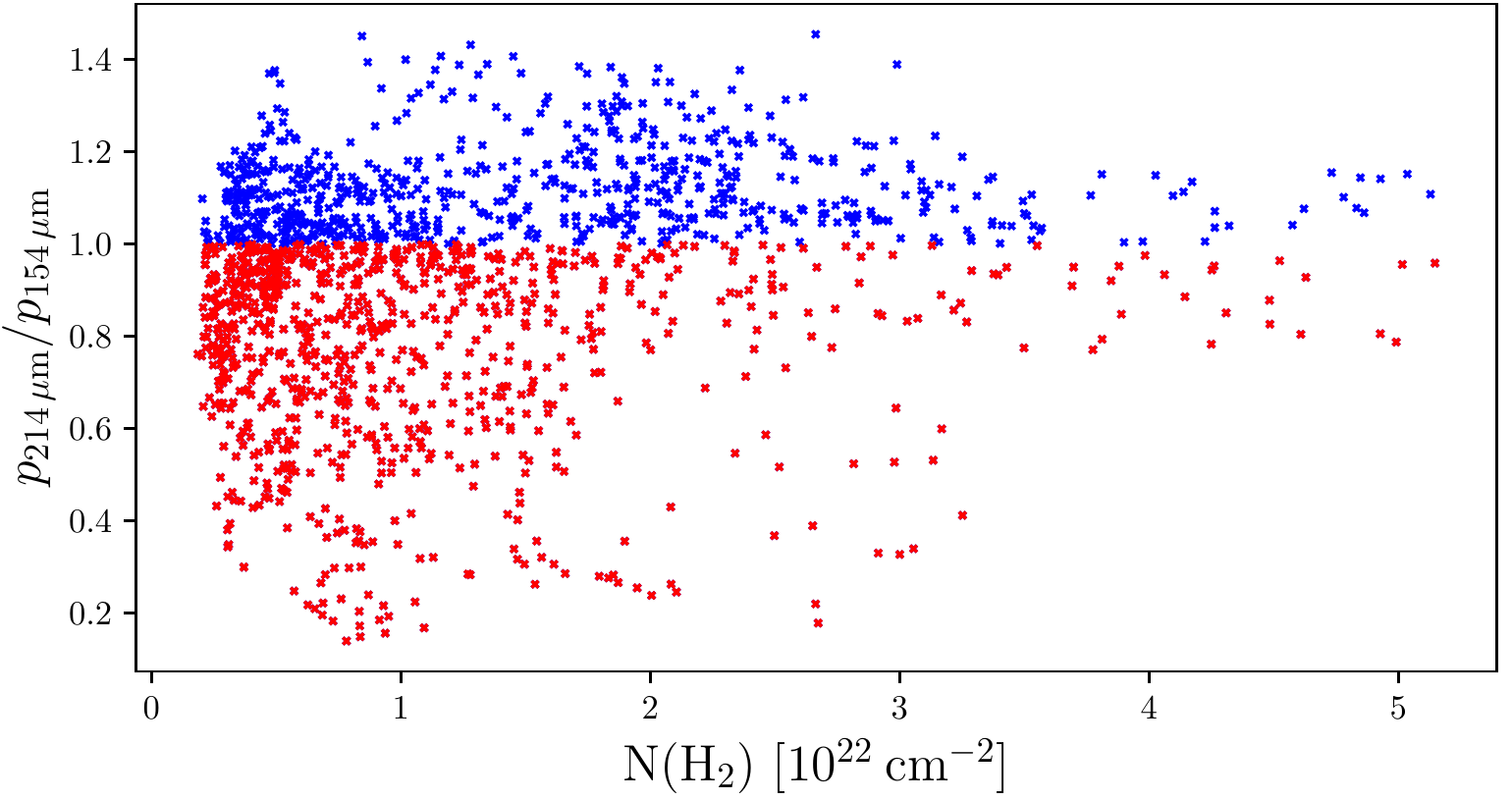}
\includegraphics[width=\hsize]{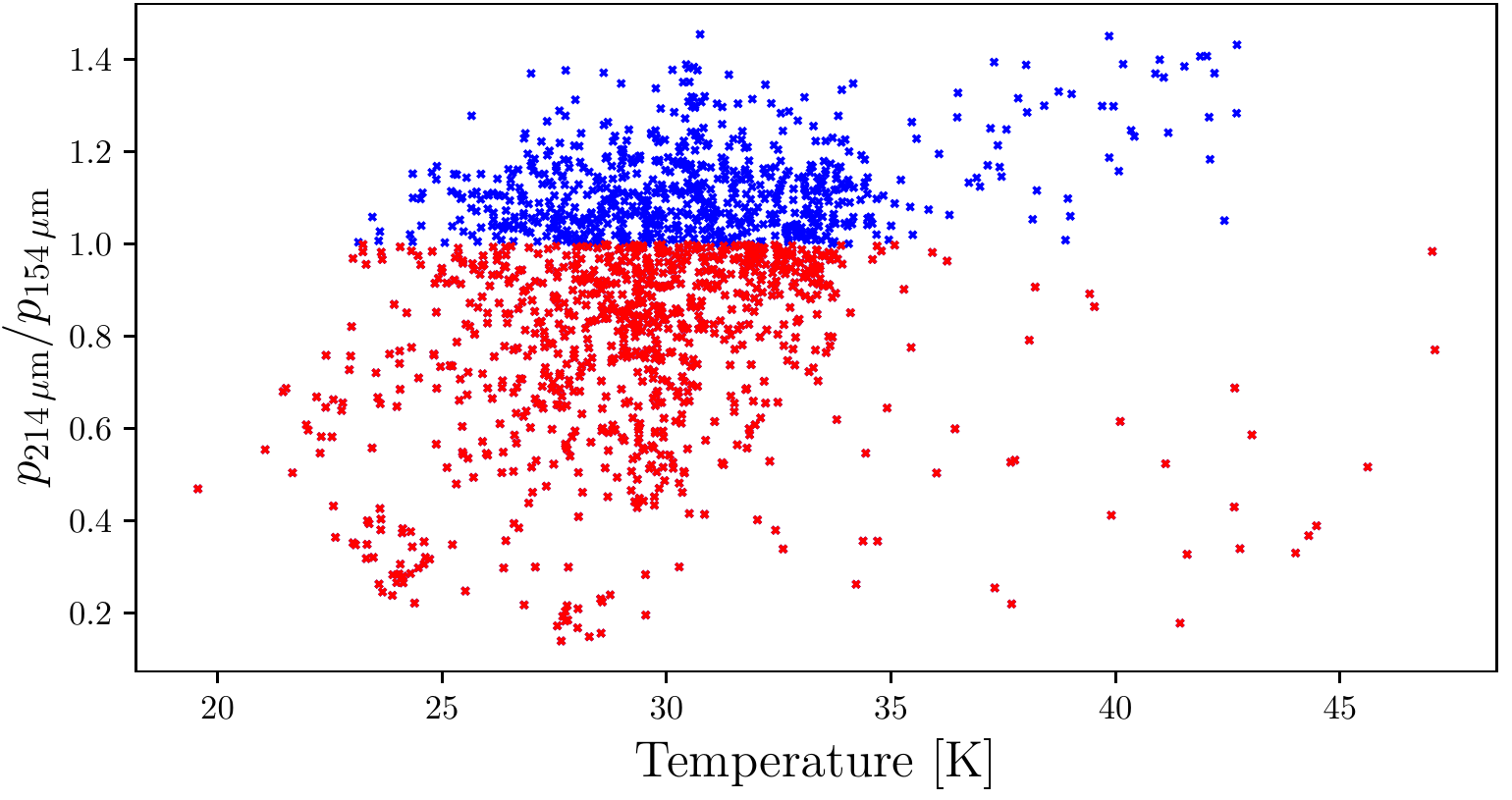}
\caption{Polarization spectrum ($p_{\mathrm{214\,\mu m}} / p_{\mathrm{154\,\mu m}})$ versus column density (top) and temperature (bottom). Red (blue) crosses indicate a polarization spectrum smaller than 1 (larger than 1).}
\label{Figure:Polarization_spectrum_vs_column_density_temperature}
\end{figure}

The spatially resolved map of $p_{\mathrm{214\,\mu m}} / p_{\mathrm{154\,\mu m}}$ is shown in Fig. \ref{Figure:Map_polarization_spectrum}.
In the southern and eastern part of OMC-3, the polarization spectrum is smaller than 1, while in the central and northern part the spectrum is mostly greater than 1. The ratio of
$p_{\mathrm{214\,\mu m}} / p_{\mathrm{154\,\mu m}}$ versus column density and temperature is shown in Fig \ref{Figure:Polarization_spectrum_vs_column_density_temperature} (top) and Fig. \ref{Figure:Polarization_spectrum_vs_column_density_temperature} (bottom), respectively. In our maps, we find a slightly larger number of data points with a negative slope (878) than with a positive slope (709). The mean polarization slope is slightly negative $\overline{ (p_{\mathrm{214\,\mu m}} / p_{\mathrm{154\,\mu m}})}$ = 0.93 $\pm$ 0.24, indicating a relatively flat slope of the polarization spectrum.        
We do not find a clear correlation between polarization spectrum and column density. However, it does appear that the polarization spectrum is particularly flat ($\sim$ 1) at higher column densities (see Fig. \ref{Figure:Polarization_spectrum_vs_column_density_temperature} top). The polarization spectrum is smaller than 1 for low ($\lesssim$ 25\,K) and high ($\gtrsim$ 42\,K) temperatures.  However, in these cases the sample size is small if compared to the total number of data points. Therefore, no significant conclusion about the connection between the polarization ratio and the derived temperature is possible.\\
In contrast, \citet{Michail2021} find a positive correlation between the slope of the polarization spectrum and the temperature of OMC-1. However, they report no significant correlation between slope and column density. In contrast, \citet{Santos2019} report a clear correlation between polarization spectrum and column density and temperature in $\rho$ Oph A. 

\section{Magnetic field of OMC-3 derived from observations in different wavelength ranges} \label{Section:Magnetic_field_diff_wavelengths}
\begin{figure*}
\includegraphics[width=\textwidth]{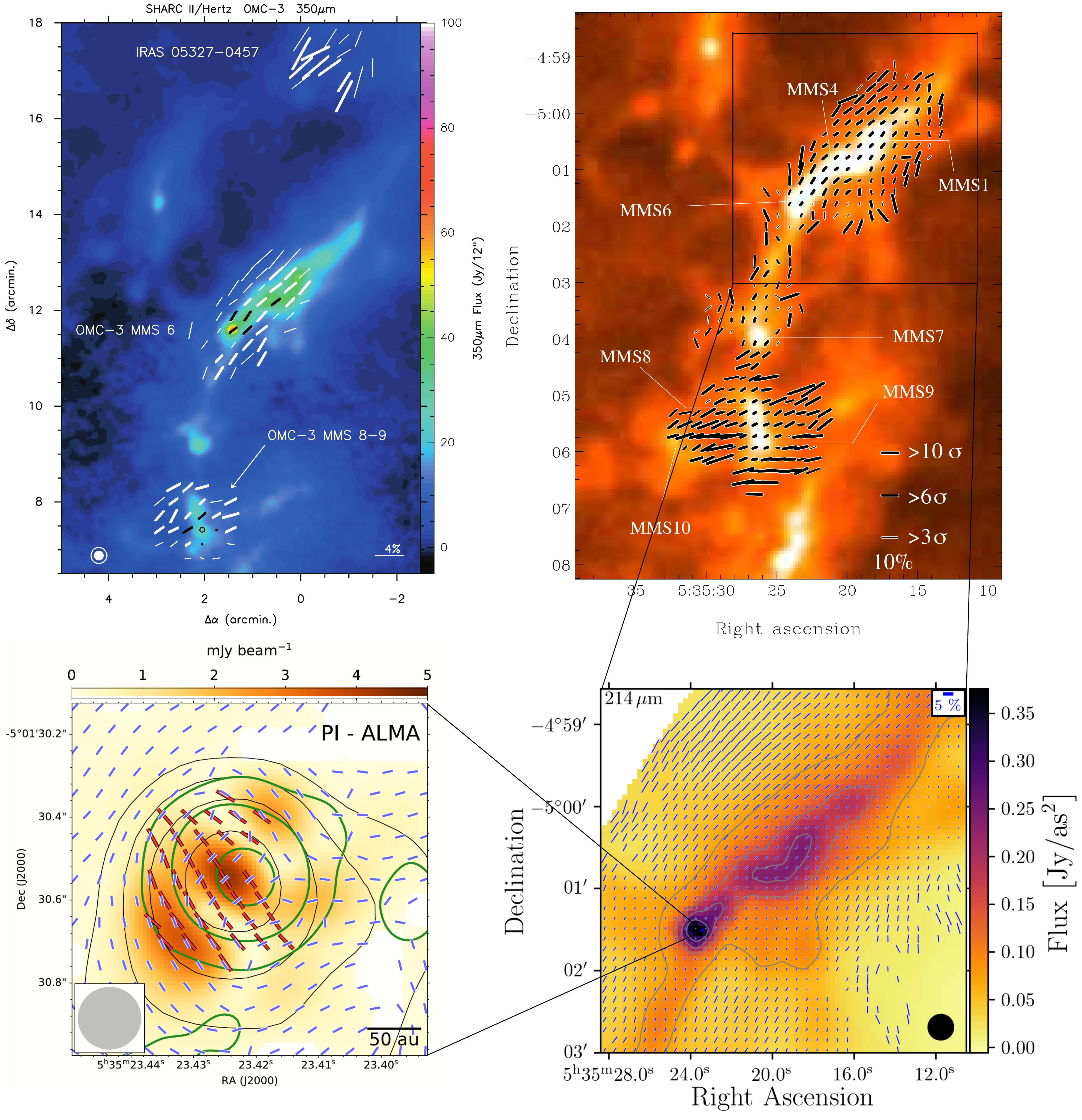}
\caption{Multiwavelength, multiscale polarization maps of OMC-3. Top left: Intensity and polarization map obtained
with SHARC II and Hertz at 350\,$\mu$m, respectively. Overlaid are polarization vectors in white and black \citep[][\copyright AAS. Reproduced with permission]{Houde2004}. Top right: Total intensity, observed at 850\,$\mu$m with SCUBA, is shown with overlaid polarization vectors in blue \citep[][\copyright AAS. Reproduced with permission]{Matthews2001}. Bottom left: Total intensity, observed at 1.2\,mm with ALMA, is shown with overlaid polarization vectors in blue. Red polarization vectors are oberserved with JVLA at 9\,mm \citep[][\copyright AAS. Reproduced with permission]{Liu2020}. Bottom right: Total intensity, observed at 214\,$\mu$m with SOFIA/HAWC+, is shown with overlaid polarization vectors in blue.} \label{Figure:OMC3_at_different_wavelengths}
\end{figure*}

OMC-3 is a well studied object with polarimetic observations ranging from the far-infrared to submm and mm. In the following, we compare these observations to the SOFIA/HAWC+ observation at 154 and 214\,$\mu$m (see Fig. \ref{Figure:OMC3_at_different_wavelengths}).

\begin{itemize}
\item[] 350\,$\mu$m: \citet{Houde2004} observed OMC-3 with SHARC II/Hertz at 350\,$\mu$m. These authors reported a mean polarization angle of $\bar{\sigma}$ = -43$^\circ\pm9^\circ$ for OMC-3 MM1-6. If we limit our observations to the same region, we obtain \mbox{$\bar{\sigma}$ = -34$^\circ\pm12^\circ$} and $\bar{\sigma}$ = -34$^\circ\pm10^\circ$ at 154\,$\mu$m and 214\,$\mu$m, respectively. The Stokes averaged polarization degree at 350\,$\mu$m amounts to $\overline{p_\mathrm{350\,\mu m}}$ = 1.55$\%$ $\pm$ 0.12$\%$. For the same region we obtain $\overline{p_\mathrm{154\,\mu m}}$ = 3.2$\%$ $\pm$ 1.71$\%$ and \mbox{$\overline{p_\mathrm{214\,\mu m}}$ = 2.86$\%$ $\pm$ 1.35$\%$.} While the polarization angles are well aligned at 154\,$\mu$m, 214\,$\mu$m, and 350\,$\mu$m, the polarization degree is 1-2$\%$ lower at 350\,$\mu$m. The rather small standard deviation of the polarization degree at 350\,$\mu$m indicates that here the polarization hole is not as prominent as it is at 154 and 214\,$\mu$m. \\

\item[] 850\,$\mu$m: \citet{Matthews2001} observed OMC-3 with JCMT/SCUBA at 850\,$\mu$m. 
The reported polarization angles are well aligned with our results. 
The mean polarization degree at 850\,$\mu$m is 5.0\,$\%$, including the observation at the southern regions (MMS7-10, see Fig. \ref{Figure:OMC3_at_different_wavelengths}, top-right). This mean polarization degree is similar to our results, 4.8$\pm$2.7$\%$ and 3.8$\pm$2.0$\%$ for 154 and 214$\,\mu$m, respectively. The 850\,$\mu$m observation shows a polarization hole for OMC-3 as well. \\

\item[] 1.2\,mm $\&$ 9\,mm: \citet{Liu2020} observed OMC-3 with ALMA/JVLA at 1.2 and 9\,mm. These high-resolution observations show that the polarization angles have a more complex pattern at small scales than they have at larger scales. Since SOFIA/HAWC+ does not allow resolving these structures, we do not compare the polarization degrees. 

\end{itemize}
The magnetic field structure, which is derived from the SOFIA/HAWC+ observations, is consistent with the magnetic field, which was reported from previous polarimetric observations in the far infrared and submm wavelength range. While the magnetic field appears uniform at larger scales, it shows a a greater level of complexity on small scales. 


\section{Conclusions} \label{Section_Conclusion}
We investigated the magnetic field of OMC-3 based on polarimetric observations with SOFIA/HAWC+ at 154\,$\mu$m and 214\,$\mu$m.

\begin{itemize}
\itemsep 10pt

\item[1.] The polarization maps of OMC-3 at 154\,$\mu$m (band D) and 214\,$\mu$m (band E) show a uniform pattern, parallel to the filament for both wavelengths. The mean polarization angles are \mbox{$\overline{\theta_\mathrm{154\,\mu m}} = -32.6^\circ$ $\pm$ 14.5$^\circ$} and $\overline{\theta_\mathrm{214\,\mu m}} = -24.1^\circ$ $\pm$ 20.4$^\circ$ for 154\,$\mu$m and 214\,$\mu$m, respectively. These results are consistent with previous polarimetric observations of OMC-3 in the far-infrared and submm wavelength range \citep[][]{Matthews2001, Houde2004}.

\item[2.] The mean polarization degree amounts to $\overline{p_\mathrm{154\,\mu m}} = 4.8\,\%$ $\pm$ 2.7$\,\%$ and $\overline{p_\mathrm{214\,\mu m}} = 3.8\,\%$ $\pm$ 2.0$\,\%$ at 154\,$\mu$m and 214\,$\mu$m, respectively. The polarization degree decreases for both wavelengths toward regions with increased column density. An unequivocal explanation for the occurence of this "polarization hole" could not be found. However,  the "optical depth effect", namely, the superposition of polarized emission and dichroic extinction seems to be of importance in the innermost densest regions, consistent with the observed  90$^\circ$ flip of the polarization vectors that has been observed using ALMA and JVLA. 

\item[3.] The magnetic field of OMC-3 is uniform and perpendicular to the filament for both wavelengths. We calculated a magnetic field strength of 202\,$\mu$G  at 154\,$\mu$m and 261\,$\mu$G at 214\,$\mu$m.

\item[4.] We do not find a general correlation between the polarization spectrum ($p_{\mathrm{214\,\mu m}} / p_{\mathrm{154\,\mu m}})$ and cloud properties, that is, column density $N(H_2)$ and temperature $T$. These results are in contrast to previous studies of similar objects \citep[][]{Santos2019, Michail2021}.

\item[5.] The large-scale magnetic field structure and strength are consistently derived from observations that cover a  wide range of wavelengths, that is, the far infrared to submm. On small scales, the magnetic field appears to be more complex. 

\end{itemize}
Using SOFIA/HAWC+ we obtained new multiwavelength polarimetric observations of the filamentary structure OMC-3 at 154\,$\mu$m and 214\,$\mu$m. These observations reveal a uniform magnetic field, which is oriented perpendicular to the filament. These findings are in good agreement with previous observations at similar scale. No correlation between the polarization spectrum and cloud properties of OMC-3 has been found.

\section*{Acknowledgements}
We thank Stefan Heese for data aquisition. This paper is based on observations made with the NASA/DLR Stratospheric Observatory for Infrared Astronomy (SOFIA). SOFIA is jointly operated by the Universities Space Research Association, Inc. (USRA), under NASA contract NNA17BF53C, and the Deutsches SOFIA Institut (DSI) under DLR contract 50 OK 0901 to the University of Stuttgart.  Acknowledgement: N.Z. and S.W. acknowledge the support by the DLR/BMBF grant 50OR1910. The LIC code is ported from publically-available IDL source by Diego Falceta-Gon\c{c}alves.

\bibliographystyle{aa} 
\bibliography{lit}

\onecolumn
\begin{appendix}

\section{Polarization map of OMC-3}
\begin{figure*} [h]
   \centering
   \includegraphics[width=\hsize]{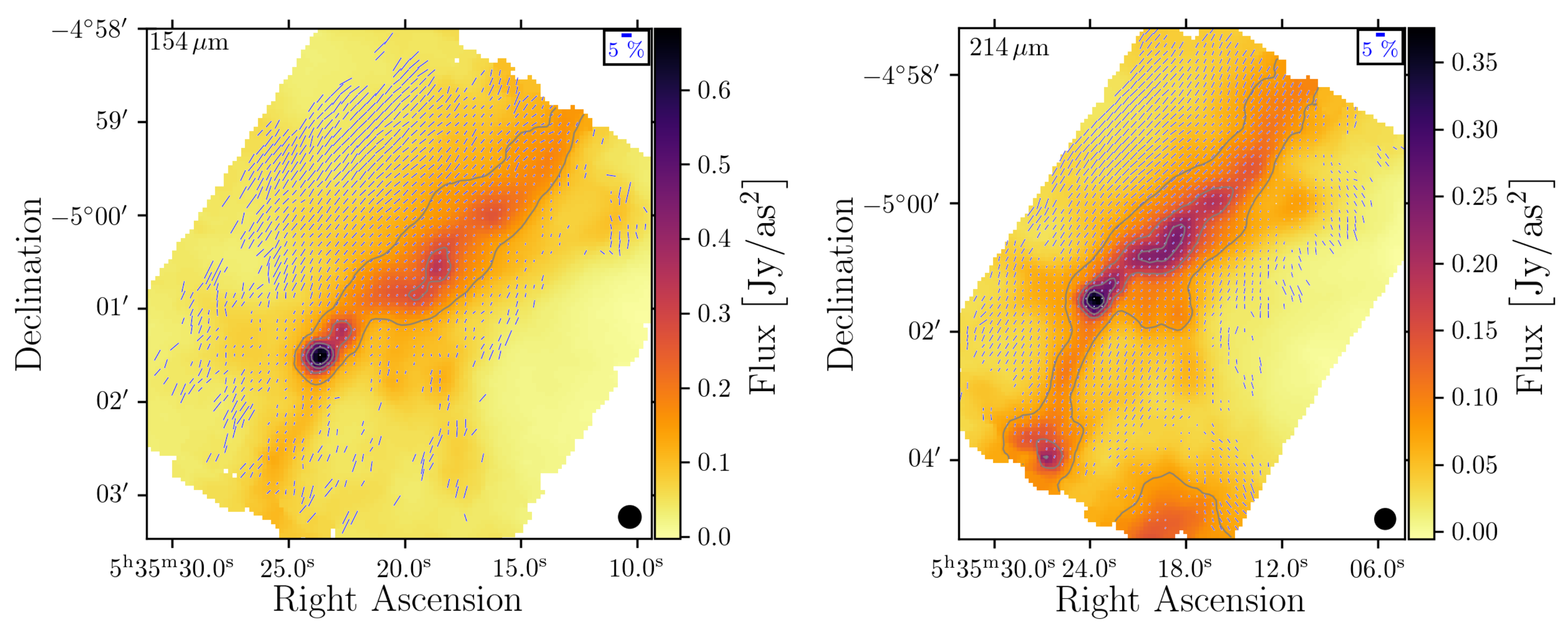}
      \caption{Complete SOFIA/HAWC+ band D (154\,$\mu$m, left) and E (214\,$\mu$m, right) polarization maps of OMC-3. The total intensity is shown with overlaid polarization vectors in blue. The length of the vectors is proportional to the polarization degree and the direction gives the orientation of the linear polarization. The isocontour lines mark 20, 40, 60, and 80$\%$ of the maximum intensity. According to criteria (\ref{Formula:Requirement1}) and (\ref{Formula:Requirement2}) only vectors with $I > 100\, \sigma_I$ and \mbox{$p > 3\, \sigma_p$} are considered (see Sect. \ref{Section:Data_Aquisition}). The beam size of 13.6$''$ for band D and 18.2$''$ for band E (defined by the FWHM) are indicated on the lower-right.
              }
         \label{Polarization_Map_OMC3_Complete}
\end{figure*}

\newpage
\section{Optical depth map of OMC-3 at 154, 160, 214, and 850\,$\mu$m}

\begin{figure} [h]
   \centering
   \includegraphics[width=\hsize]{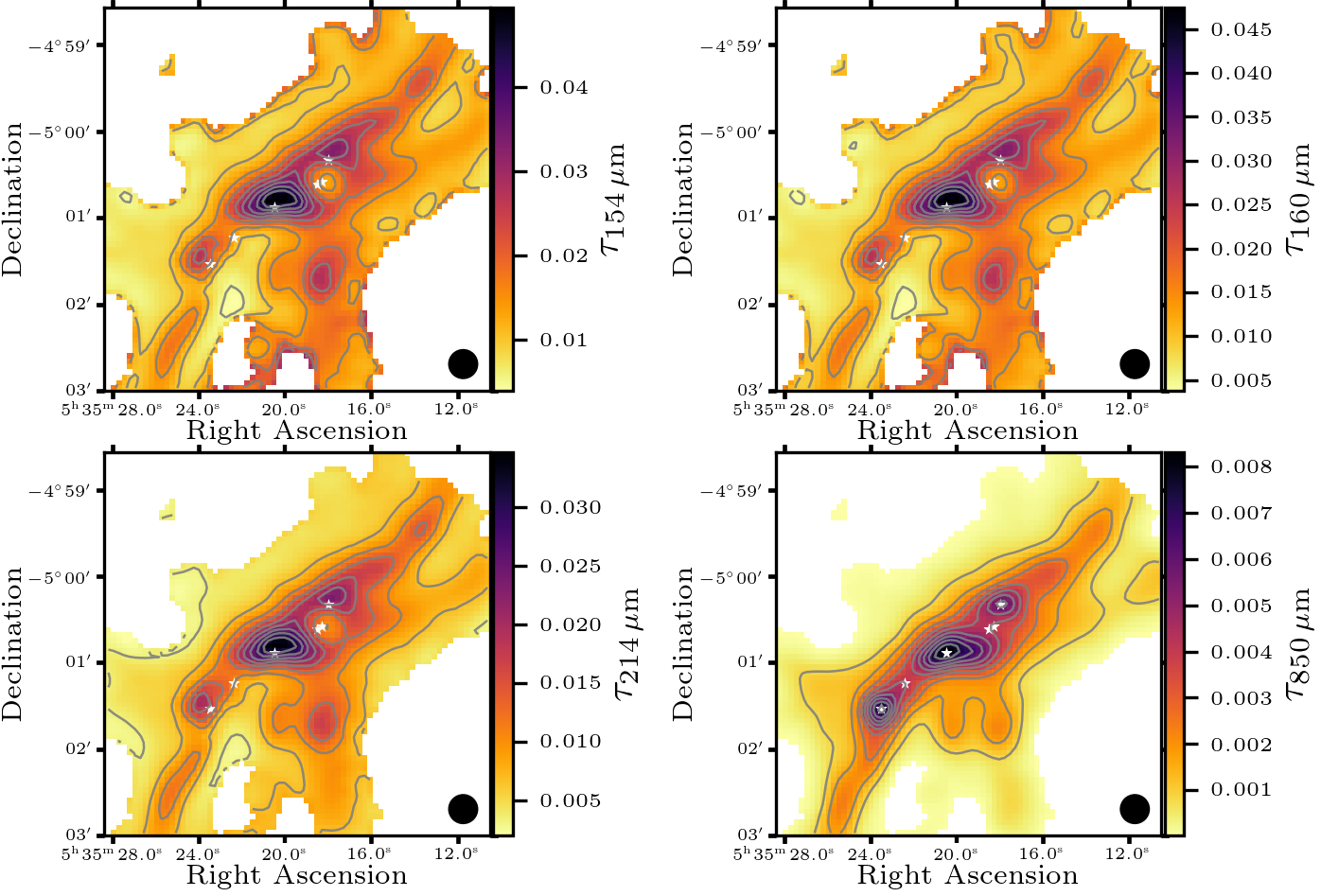}
      \caption{Maps of optical depth at wavelengths of 154\,$\mu$m (top left), 160\,$\mu$m (top right), 214\,$\mu$m (bottom left), and 850\,$\mu$m (bottom right). The contour lines mark 10, 20, 30, 40, 50, 60, 70, 80, and 90$\,\%$ of the maximum optical depth for each wavelength. The white asterisk symbols mark known stellar sources \citep{Chini1997}. The beam size of 18.2$''$ (band E)  is indicated on the lower-right of each figure.
              }
         \label{Figure:Maps_Optical_Depth}
\end{figure}

\end{appendix}

\end{document}